\documentclass[chicago]{emulateapj}
\usepackage{CJK}
\usepackage{natbib}
\usepackage{apjfonts}
\begin{document}

\begin{CJK*}{UTF8}{gbsn}
\title{GAMMA-RAY
BURST PROMPT EMISSION LIGHT CURVES AND POWER DENSITY SPECTRA
IN THE ICMART MODEL}
\author{Bo Zhang (张博)\altaffilmark{1,2,3} and Bing Zhang (张冰)\altaffilmark{2}}
\altaffiltext{1}{School of Astronomy and Space Science, Nanjing University, Nanjing 210093,
China; bozhang@physics.unlv.edu}
\altaffiltext{2}{Department of Physics and Astronomy, University of Nevada Las
Vegas, Las Vegas, NV 89154, USA}
\altaffiltext{3}{Purple Mountain Observatory, Chinese Academy of Sciences, Nanjing 210003, China}

\begin{abstract}
In this paper, we simulate the prompt emission light curves of
gamma-ray bursts (GRBs) within the framework of the 
Internal-Collision-induced MAgnetic Reconnection and Turbulence
(ICMART) model. This model applies to GRBs with a 
moderately high magnetization parameter $\sigma$ in the emission region.
We show that this model can produce highly variable light curves 
with both fast and slow components. The rapid variability is
caused by many locally Doppler-boosted mini-emitters due to turbulent
magnetic reconnection in a moderately high $\sigma$ flow. The runaway 
growth and subsequent depletion of these mini-emitters as a function of time
define a broad slow component for each ICMART event. A GRB light curve
is usually composed of multiple ICMART events that are fundamentally
driven by the erratic GRB central engine activity. Allowing variations
of the model parameters, one is able to reproduce a variety of light
curves and the power density spectra as observed. 
\end{abstract}
\maketitle
\noindent

\section{Introduction\label{sec:intro}}
A gamma-ray burst (GRB) event comprises two phases, prompt emission 
and afterglow. The prompt $\gamma$-ray emission is usually highly 
variable, with many pulses overlapping within a short duration 
(Fishman \& Meegan 1995).
The power density spectra (PDSs) of the light curves are typically a
power law with a possible turnover at high frequencies (Beloborodov
et al. 2000). The light curves may be decomposed as the superposition
of an underlying slow component and a more rapid fast component 
(Gao et al. 2012). The fast component tends to be more significant in
high energies, and becomes less significant at lower
frequencies (Vetere et al. 2006).

It has been shown that the external shock model has difficulty 
producing GRB variability while maintaining a high radiative efficiency
(Sari \& Piran 1997; cf. Dermer \& Mitman 1999).  The detection of
the steep decay phase following GRB prompt emission (Tagliaferri et
al. 2005) suggests that the prompt emission region is detached from
the afterglow emission region (Zhang et al. 2006). This nails down
the internal origin of GRB prompt emission for the majority of GRBs.

For an internal origin of GRB prompt emission, the variability 
is usually attributed to the erratic activity of the central engine 
(e.g., Rees \& M\'esz\'aros 1994; Kobayashi et al. 1997). It is 
envisaged that the ejecta launched from the central engine is composed
of multiple shells with variable bulk Lorentz factors.
Faster late shells catch up and collide with slower early shells. Part 
of the kinetic energy of the ejecta is converted to energy of 
non-thermal particles in these internal shocks, a fraction of which
is released as the observed non-thermal radiation. In this model, 
different variability timescales are related to the angular spreading 
time of colliding shells at different internal shock radii. In order
to account for superposed slow and fast variability components, one has
to assume that the central engine itself carries these two variability
components in the time history of jet launching (Hasco\"et et al. 2012),
whose physical origin is unclear. The internal shock model 
also suffers a list of criticisms (e.g., Zhang \& Yan 2011 for a review), 
including low radiation efficiency (e.g., Kumar 1999;
Panaitescu et al. 1999), fast cooling (Ghisellini et al. 2000; Kumar
\& McMahon 2008),\footnote{This problem is recently alleviated by
Uhm \& Zhang (2013), who showed that by introducing magnetic field 
decay as the outflow streams outward, the fast cooling spectrum can
be harder than the traditional $F_\nu \propto \nu^{-1.5}$ spectrum.
However, a requirement is that the emission region has to be large
where the magnetic field is weak. This corresponds to an unconventional
internal shock radius, but is consistent with the ICMART model.}, 
particle number excess (Daigne \& Mochkovitch 1998; 
Shen \& Zhang 2009), 
inconsistency with some empirical relations
(Amati et al. 2002; Zhang \& M\'esz\'aros 2002; Liang et al. 2010), 
and overpredicting the brightness of the photosphere emission 
component (Daigne \& Mochkovitch 2002; Zhang \& Pe'er 2009).

Alternatively, the GRB variability can be interpreted as
locally Doppler-boosted emission in a relativistic bulk flow, such
as relativistic mini-jets (Lyutikov \& Blandford 2003; Yamazaki et al. 2004) 
or relativistic turbulence (Narayan \& Kumar 2009; Kumar \&
Narayan 2009; Lazar et al. 2009) in a bulk relativistic ejecta. Some
criticisms have been raised to these models. For example,
relativistic turbulence damps quickly so that the emission from the
turbulence cannot be sustained (Zrake \& MacFadyen 2012). The 
simulated light curves are composed of well-separated sharp pulses 
without an underlying slow component (Narayan \& Kumar 2009; 
Lazar et al. 2009). Also the pulse was calculated to have a 
symmetric shape for the turbulence model (Lazar et al. 2009), 
which is in contradiction with the data. 

Recently, Zhang \& Yan (2011, hereafter ZY11) proposed an
Internal-Collision-induced MAgnetic Reconnection and Turbulence
(ICMART) model to explain prompt emission of GRBs.  
Like the traditional internal shock scheme, the ICMART model envisages 
internal interactions of shells within the ejecta wind. The main 
difference is that the ejecta is Poynting flux dominated, with the
magnetization parameter $\sigma \equiv F_{\rm P}/F_{\rm m} > 1$ in the
collision region, where $F_{\rm P}$ and $F_{\rm m}$ are Poynting
flux and matter flux, respectively. This was motivated by the
non-detection of a bright photosphere thermal component in GRB 080916C
(Zhang \& Pe'er 2009) and most other Large Area Telescope GRBs (Zhang et al. 2011). 
For a helical magnetic field structure, the 
initial collisions only serve to distort the magnetic field 
configurations. As multiple collisions proceed, the field 
configurations would be distorted to a critical point when a 
cascade of reconnection and turbulence occurs. Charged particles 
can be accelerated in these reconnection regions, leading to intense 
gamma-ray radiation. Within this model, a GRB light curve is supposed 
to have two variability components: a broad (slow) component that 
tracks central engine activity, and an erratic (fast) component with 
multiple sharp pulses superposed on the slow component, which is 
related to numerous reconnection sites during the ICMART event. 

In this paper, we simulate GRB light curves and their corresponding
PDSs within the framework of 
the ICMART model. In Section 2 we describe the basic model and the
simulation method. The simulation results are presented in Section
3. Section 4 summarizes the findings with some discussion.

\section{Basic Scheme and Simulation Methods\label{sec:basic}}
We first summarize the basic ideas of the ICMART model (ZY11).
Magnetized shells with initial $\sigma > 1$ are envisaged to
collide, leading to distortion of magnetic field lines until a
threshold is reached and a runaway magnetic dissipation is
triggered. During such an ``avalanche''-like reconnection/turbulence 
cascade, it is envisaged that fast reconnection seeds in the 
moderately high $\sigma$ regime would inject moderately
relativistic outflows in the emission regions
(ZY11; Lyubarsky 2005), which would
excite relativistic turbulence. The turbulence would facilitate more
reconnection events, which trigger further turbulence. The magnetic
energy is converted to particle energy and efficient radiation. 
During the growth of the reconnection/turbulence cascade, the number 
of reconnection sites as observed at any instant increases rapidly
with time, so that multiple mini-emitters contribute
simultaneously to the observed gamma-ray emission. Rapid evolution
of individual reconnection sites leads to rapid variability of the
observed GRB light curves. The cascade stops as $\sigma$ drops around
or below unity when most magnetic energy is converted into radiation
or kinetic energy. During the growth of an ICMART event, turbulence
is not quickly damped due to the continuous injection of particle
energy from the reconnection events, which continuously drives
turbulence.

With these preparations, we can model the light curve of a GRB
within the framework of the ICMART model. Lacking full numerical
simulations of magnetic turbulence and reconnection, in this paper
we perform a Monte Carlo simulation based on some simplest
assumptions. We define each reconnection 
event as a fundamental mini-emitter, which carries a local Lorentz
boost with respect to the bulk of the emission outflow. Each 
reconnection event can be modeled as a pulse, which can be bright
and spiky if the mini-emitter beams toward the observer, but dim 
and broad if the mini-emitter beams away from the
observer's direction. The observed light curve is the superposition 
of the emission from all these mini-emitters. For simplicity, we 
assume that the characteristic brightness (peak luminosity) of each
reconnection event in the rest frame of the reconnection outflow is
the same. We also take the shape of each pulse as a Gaussian form
for simplicity (e.g., Narayan \& Kumar 2009; Lazar et al. 2009).
Our goal is to try to simulate the superposed slow and
fast components, and the precise shape of each pulse does not
matter too much. In any case, we note that the shape of a 
spike within the ICMART model is mainly defined by the time
history of each reconnecting mini-jet rather than the time history 
of an ideal eddy, so the pulse profile may not necessarily be 
symmetric with peak time. This is different from the previous models 
(Narayan \& Kumar 2009; Lazar et al. 2009) that invoke relativistic
turbulence.\footnote{Even for turbulence, the previous simulations
invoked a circular eddy, while MHD turbulence eddies are highly
distorted, especially in small scales (Goldreich \& Sridhar 1995;
Cho, Lazarian \& Vishniac 2003). This would reduce the symmetry of the pulse
shapes in the turbulence models. As a result, we consider the
symmetry issue raised by Lazar et al. (2009) would not be relevant
for the ICMART model.} More importantly, the shape of a broad
pulse in the model is asymmetric: the rising portion is defined by 
the timescale of the reconnection-turbulence cascade process, while 
the decay portion is controlled by high-latitude emission after the
ICMART cascade ceases.

There are three rest frames in this model: the first is the rest
frame of the mini-jet, i.e. the outflow of the individual reconnection
event. These mini-jets are moving with a relative Lorentz factor
$\gamma$ with respect to the jet bulk. We denote parameters in this
frame as ($''$). The second frame is the rest frame of the jet bulk,
which moves with a Lorentz factor $\Gamma$ with respect to the
central engine. We denote parameters in this frame as ($'$). The 
third one is the rest frame of the observer
(with the cosmological expansion effect ignored). The quantities
within these three frames are connected through two Doppler factors,
i.e.,
\begin{equation}
 {\cal D}_1 = \left[\Gamma\left(1- \beta_{\rm{bulk}} \cos \theta
\right)\right]^{-1}
\end{equation}
and 
\begin{equation}
 {\cal D}_2 = \left[\gamma\left(1- \beta \cos \phi
\right)\right]^{-1},
\end{equation}
where $\beta_{\rm{bulk}}$ and $\beta$ are the corresponding dimensionless
velocities with respect to $\Gamma$ and $\gamma$, respectively, 
$\theta$ is the latitude of the mini-jet with respect to
the line of sight (i.e. the angle between the line of sight and the
radial direction of the bulk ejecta at the location of the mini-jet),
and $\phi$ is the angle between the mini-jet direction and radial
direction of the ejecta bulk within the comoving frame of the ejecta
bulk. 

Each reconnection event is supposed to give rise to a single
pulse in the GRB light curve. Since several reconnection events may
occur simultaneously, some pulses can superpose with each other.  
For a naive Sweet-Parker reconnection,\footnote{In reality, 
the Sweet-Parker prescription is
too simple. Rapid reconnection is achieved in an X-point geometry
with turbulence playing an essential role (e.g. Lazarian \&
Vishniac 1999). The simple treatment here only offers an order
of magnitude estimate.} one has
(e.g., see Zweibel \& Yamada 2009 and references therein)
\begin{equation}
v'_{\rm{in}} L' = v'_{\rm{out}} r',
\end{equation}
where $v'_{\rm{in}}$ is the inflow velocity 
of the reconnection layers, $v'_{\rm{out}}$ is the outflow velocity,
and $r'$ and $L'$ are the width and length of the reconnection layer, 
respectively. Reconnection physics demands $v'_{\rm{in}} \ll v'_{\rm{out}}$,
so that $r' \ll L'$. On the other hand, what defines the duration 
of the reconnection event is the thickness of the bunch of magnetic 
field lines that continuously approach each other, and we assume
that it is also of the order of $L'$. As a result, in
the bulk comoving frame (the $'$ frame), the duration
of each pulse can be approximated as $\Delta t'=
L'/v'_{\rm{in}}$. In the observer frame, this is translated to
$\Delta t = \Delta t' / {\cal D}_1$, which corresponds to 
the duration of a certain pulse in the observer frame.

For simplicity, we assume that the radiation intensity arising
from each reconnection event has the same spectral form, i.e.,
the Band function (see Band et al. 1993), in the
comoving frame of the mini-jet (the $''$ frame),
\begin{equation}
I^{''}_{\nu^{''}} \propto
\left(\frac{\nu^{''}}{\nu_0^{''}}\right)^{\alpha} \left( 1 +
\frac{\nu^{''}}{\nu_0^{''}} \right)^{\beta-\alpha}.
\end{equation}
The observed flux can be calculated as
\begin{equation}
 F_\nu = \int {\cal D}_1^3 {\cal D}_2^3 I^{''}_{\nu^{''}} 
{ d} \Omega \approx {\cal D}_1^3 {\cal D}_2^3  
I^{''}_{\nu^{''}} \frac{{r'}^{2} \cos \theta }{ D^{2}},
\end{equation}
where $D$ is the distance of the GRB to the observer.

In a high-$\sigma$ flow, $v'_{\rm{out}}$ can eventually reach a relativistic
speed (with Lorentz factor $\gamma$), and $v_{\rm{in}}^{'}$ can reach 
a maximum value of $0.1c$ (e.g. Lyubarsky 2005 and references herein). 
Therefore, $r' \sim 0.1 L'$. The Lorentz factor of the mini-jet is
related to $\sigma$ and would drop to unity when $\sigma$ drops
below unity. The detailed dependence is related to the complicated 
physics of relativistic reconnection. In this paper, we adopt
$\gamma \propto \left(1+\sigma\right)^{1/2}$ (i.e., $\gamma$ is proportional to
the relativistic Alfv\'en Lorentz factor). 
We also investigated other dependences between $\gamma$ and $(1+\sigma)$.
The general conclusions regarding how the simulated light-curve 
properties depend on various parameters are essentially similar. 
In the rest of the paper, we only
focus on the $\gamma \propto \left(1+\sigma\right)^{1/2}$ assumption.

In the simulations, we fix the Band function parameters as the 
following: $\alpha = -1$, $\beta = -3$,
and the peak frequency $\nu^{''}$ is chosen such that
$\langle {\cal D}_1 {\cal D}_2 \rangle h \nu^{''} \sim 300$ keV is satisfied, where 300 keV 
is the typical observed value of GRB spectral peak, and
$\langle{\cal D}_1 {\cal D}_2\rangle$ is the average value of the product of the 
two Doppler factors. Based on these assumptions, we calculate the 
received flux in the detector band of \textit{Swift} Burst Alert Telescope (BAT; i.e.,
15 - 150 keV).

In our Monte Carlo simulation, four random parameters have been
introduced. They are: (1) comoving length of the
reconnection region $L^{'}$, which is assumed to either have a typical 
value or have a power-law distribution 
with index $-5/3$ below a typical value; (2) the mini-jet direction 
(angle $\phi$ with respect to the bulk motion direction) 
in the bulk comoving frame, which is taken as isotropic or a
Gaussian distribution with respect to $\phi = 90^{\rm o}$
(see more discussion below); 
(3) the latitude of a mini-jet $\theta$ with respect
to the viewing direction, which is random
within the cone of the jet opening angle; and (4)
the epoch when a mini-jet occurs, which is taken to satisfy a
distribution of exponential growth with time, i.e.
$N(t) \propto 2^{t/t_0}$. 
The total number $N$ of the mini-jets is a free parameter,
which is defined by the requirement
that they dissipate most magnetic energy in the local emission 
regions, so that the local $\sigma$ is brought to below unity 
after each ICMART event.\footnote{The average $\sigma$ of the
ejecta can be still above unity, if the filling factor $f \ll 1$, 
since the majority of magnetic energy is still not
dissipated. A small $f$ seems to be required by the central
engine study of Lei et al. (2013), who obtained $\sigma$ values
greater than the measured typical Lorentz factors of GRBs
(Liang et al. 2010).}
Assuming that the magnetic energy density is roughly uniform within
the emission region, this number can be simply written as the
ratio between the total dissipated volume (i.e., total volume 
multiplied by the filling factor $f$) and the volume of the region 
affected by each reconnection event that powers a mini-jet. Within 
the $1/\Gamma$ cone, this number is
\begin{equation}
N \approx f \frac{4 \pi R^2 \frac{R}{\Gamma} \pi
\frac{1}{\Gamma^2}}{{L'}^3},
\label{eq:N}
\end{equation}
where $R$ is the radius of the emission region from the central
engine. 

Other input parameters include the radius of the emission
region $R$, the jet opening angle $\theta_j$,
the initial values of $\Gamma$,
and $\sigma$ (which defines the initial $\gamma$). For each
reconnection event, we assume that half of the dissipated magnetic 
energy is released in the form of photons, while the other half 
is deposited to the jet bulk and used to boost the kinetic energy
of the bulk.\footnote{Half of the dissipated energy is initially
deposited as heat, and then gets converted to kinetic energy due
to adiabatic expansion (Drenkhahn \& Spruit 2002).
} Therefore, $\Gamma$, $\sigma$, and $\gamma$ are all
functions of time during each ICMART event.

The exponential growth of magnetic dissipation eventually ends
when the local $\sigma$ drops around or below unity. Without numerical
simulations, it is unclear how abrupt the ending process is.
In this paper we just assume an abrupt cessation of the cascade 
process, so that the number
of new mini-jets drops to 0 after a particular time. The
observed ``tail'' emission after this epoch is therefore
contributed by the high-latitude emission from other mini-jets
not along the line of sight due to the ``curvature effect''
delay. This delay timescale is calculated as
$t_{\rm delay} = R\left(1 - \mu \right)/c$ with respect to the last
emission along the line of sight, where $\mu = \cos \theta$.
We calculate the contribution of all the mini-jets within $\theta_j=
5^\circ$. Although most of the received emission comes from the mini-jets 
within the $1/\Gamma$ cone, those mini-jets outside the $1/\Gamma$ cone make some 
contribution to the high-latitude emission. 
We calculate the delay timescale of each mini-jet, apply its
Doppler factor to calculate the amplitude and 
shape of the pulse, and superpose
these mini-jets to get the curvature tail of each ICMART event.

\section {Simulation Results}

\subsection{Sample Light Curves}
We run a series of Monte Carlo simulations to generate sample light 
curves. We first focus on the light curves for only one ICMART 
episode. The light curve of one GRB could be then modeled by
superposing multiple ICMART events.

We first take the following nominal parameters:
$R = 5 \times 10^{15}$ cm,  $L^{'} = 5 \times
10^{11}$ cm, $\Gamma_{\rm{ini}} = 200$, $\gamma_{\rm{ini}} = 3$, and
$N \approx 50,000$. Considering an exponential growth, 
i.e., that each reconnection seed would eject a bipolar outflow and would
stir up the ambient medium to trigger two reconnection events, one
may estimate the generation number of successive reconnection events,
$n \approx 14.6$, through the requirement
$N \approx \sum^n 2^n = 50,000$. The timescale for each 
generation in the bulk comoving frame may be estimated as
$L'/v'_{\rm{in}} \sim 10$ s, which corresponds to an observer frame timescale
$t_0 \sim 0.1$ s. This is the typical ``$e$-folding'' timescale. 
The total duration (rising timescale) of an ICMART event is
therefore $n$ times larger, i.e.m $t_{\rm{r}} = n t_0 \sim 1.5$ s,
which we adopt in the simulations.
We also assume that the observer's 
line of sight is along the jet axis, and we take a redshift $z=0$ for
simplicity.\footnote{Varying redshifts effectively stretches the 
light curves, and samples different
spectral segments in the intrinsic spectrum. We show below that
the light curve shape does not sensitivey depend on the spectral
regime. So the redshift factor plays a minor role in defining the
shape of light curves.} For a power-law distribution of $L'$,
in principle, $L'$ can extend to much smaller values. In our 
simulations, reconnection regions with $L' < 5 \times 10^{9}$ cm
are not considered, since the observed durations of these events
already meet the detector's variability limit.
In the following we test various factors that may affect the
shape of the light curves.

\subsubsection{Distribution of the Mini-jet Directions}

\begin{figure}
\begin{center}
\includegraphics*[width=7cm]{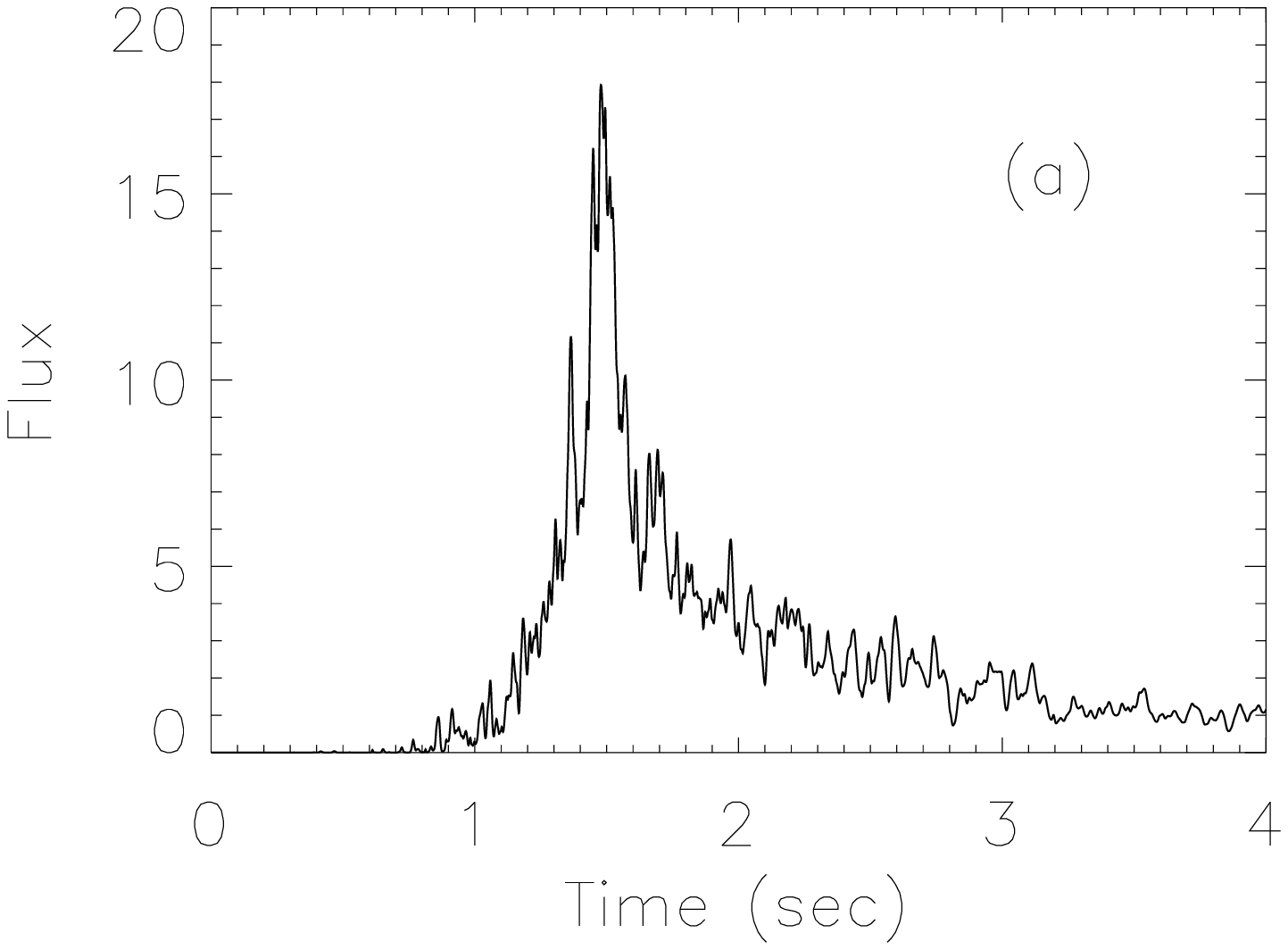}\\
\includegraphics*[width=7cm]{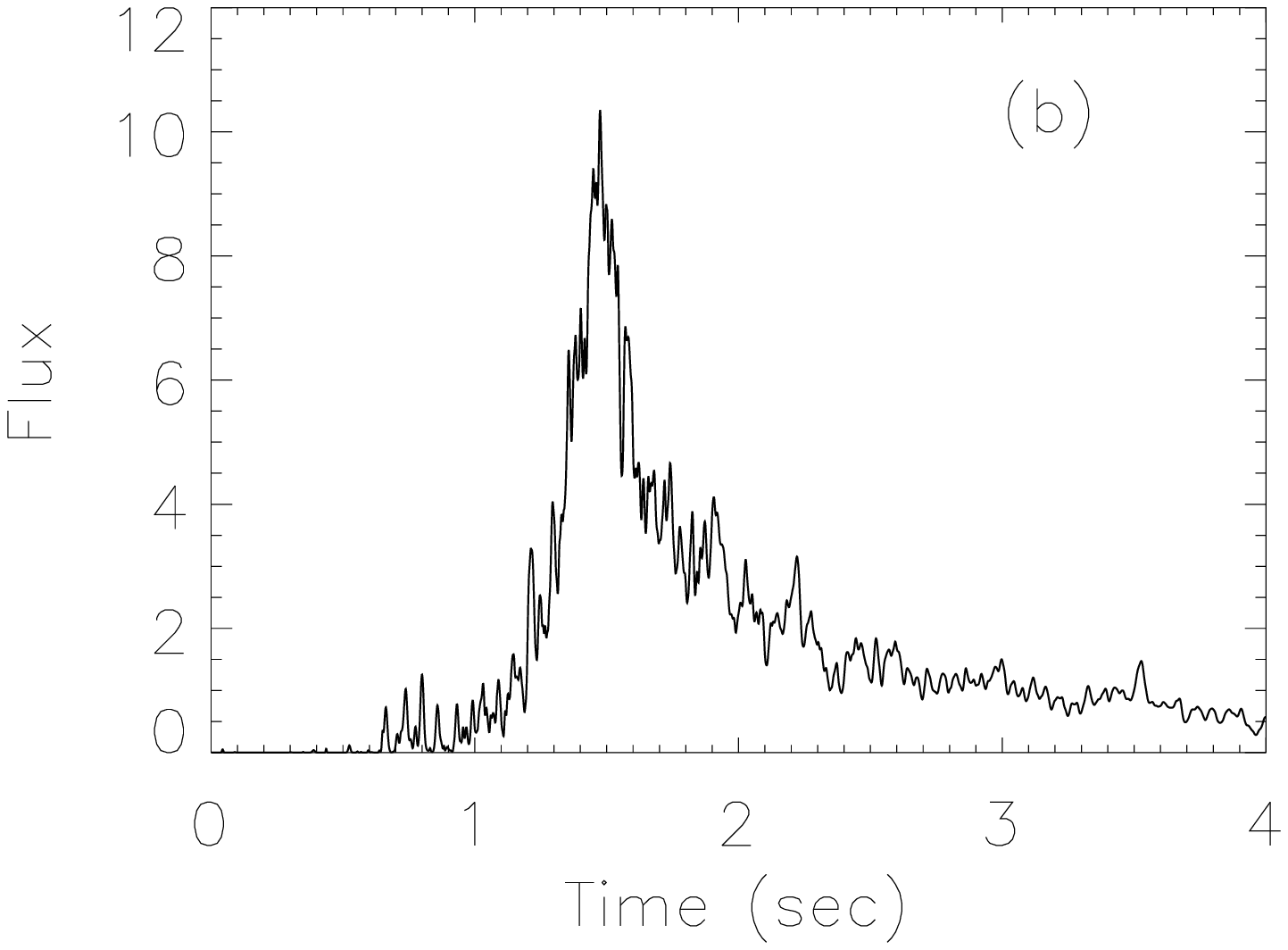}\\
\includegraphics*[width=7cm]{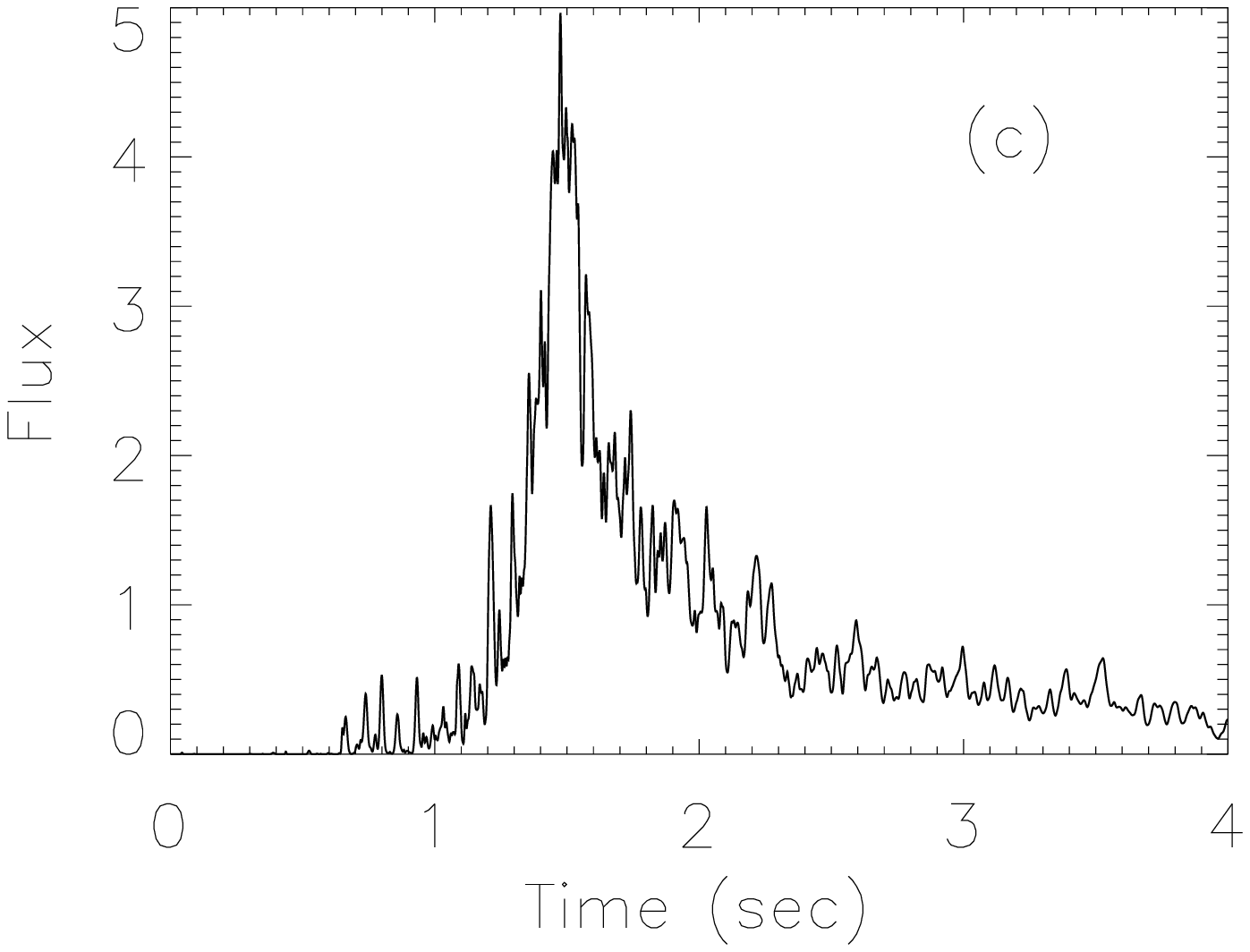}
\caption{\small{Simulated light curves of one ICMART event with
the following parameters: $R = 5\times 10^{15}$ cm,  $L^{'} = 5 \times
10^{11}$ cm, $\Gamma_{\rm{ini}} = 200$, and $\gamma_{\rm{ini}} = 3$, 
$N=50,000$, and rising time $t_r \sim 1.5$ s. The three panels 
correspond to different $\phi$ distributions: (a) isotropic;
(b) Gaussian $\phi$-distribution with a typical angle $45^\circ$
with respect to plane perpendicular to the bulk motion direction;
(c) Gaussian $\phi$-distribution with a typical angle $30^\circ$. }}\label{fig:direction}
\end{center}
\end{figure}

We first test how the simulated light curve depends on 
the unknown distribution of $\phi$ in the bulk comoving frame. 
We first assume an isotropic distribution and calculate the light 
curve. The result is shown in Figure \ref{fig:direction}(a). One can immediately 
see that the light curve has a broad component, with some spiky
small pulses superposed on top. The broad component is due to
the contributions of all the mini-jets beaming toward random
directions in the bulk motion rest frame. 
The rising of the broad pulse corresponds to the
exponential growth of the number of mini-jets, while the decay
is controlled by the high-latitude effect. 

Since an ICMART event corresponds to an event of destroying the
initial ordered magnetic field, the magnetic configurations in
the ICMART region, even near the end of the cascade, should
not be completely random. The initial magnetic field configuration
should be parallel to the ejecta plane (e.g., Spruit et al. 2001;
Zhang \& Kobayashi 2005). This is because the toroidal component
falls with radius much slower than the poloidal component. Such
a configuration should still leave an imprint on the
$\phi$ distribution. We consider a distribution of $\phi$
that has a Gaussian distribution with respect to the original
field line direction, i.e., $\phi = 90^{\rm o}$. In 
Figures \ref{fig:direction}(b) and \ref{fig:direction}(c)
we show the Gaussian angle to be $45^\circ$ and $30^\circ$,
respectively. One can see that the simulated light curves have
progressively less flux as the distribution angle becomes smaller.
This is because with a smaller distribution angle, only rare
mini-jets could beam toward the observer, which have a relatively 
lower flux (than the larger Gaussian angle distribution) 
with respect to the majority of mini-jets that beam
away from the observer and only contribute to the background.
The overall shape of the light curves does not differ significantly.

\subsubsection{Lorentz Factor Contrast}

\begin{figure}
\begin{center}
\includegraphics*[width=7cm]{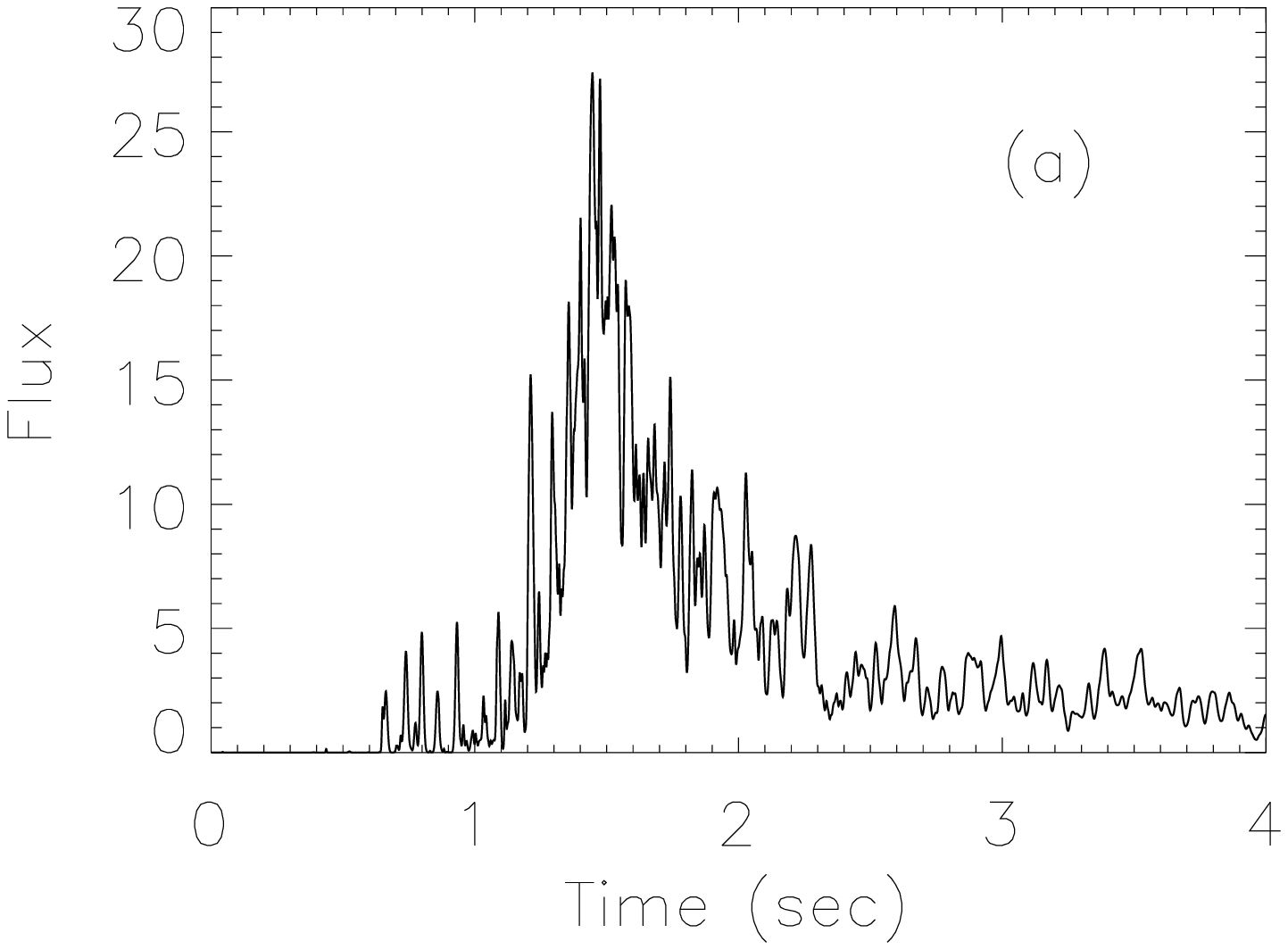}\\
\includegraphics*[width=7cm]{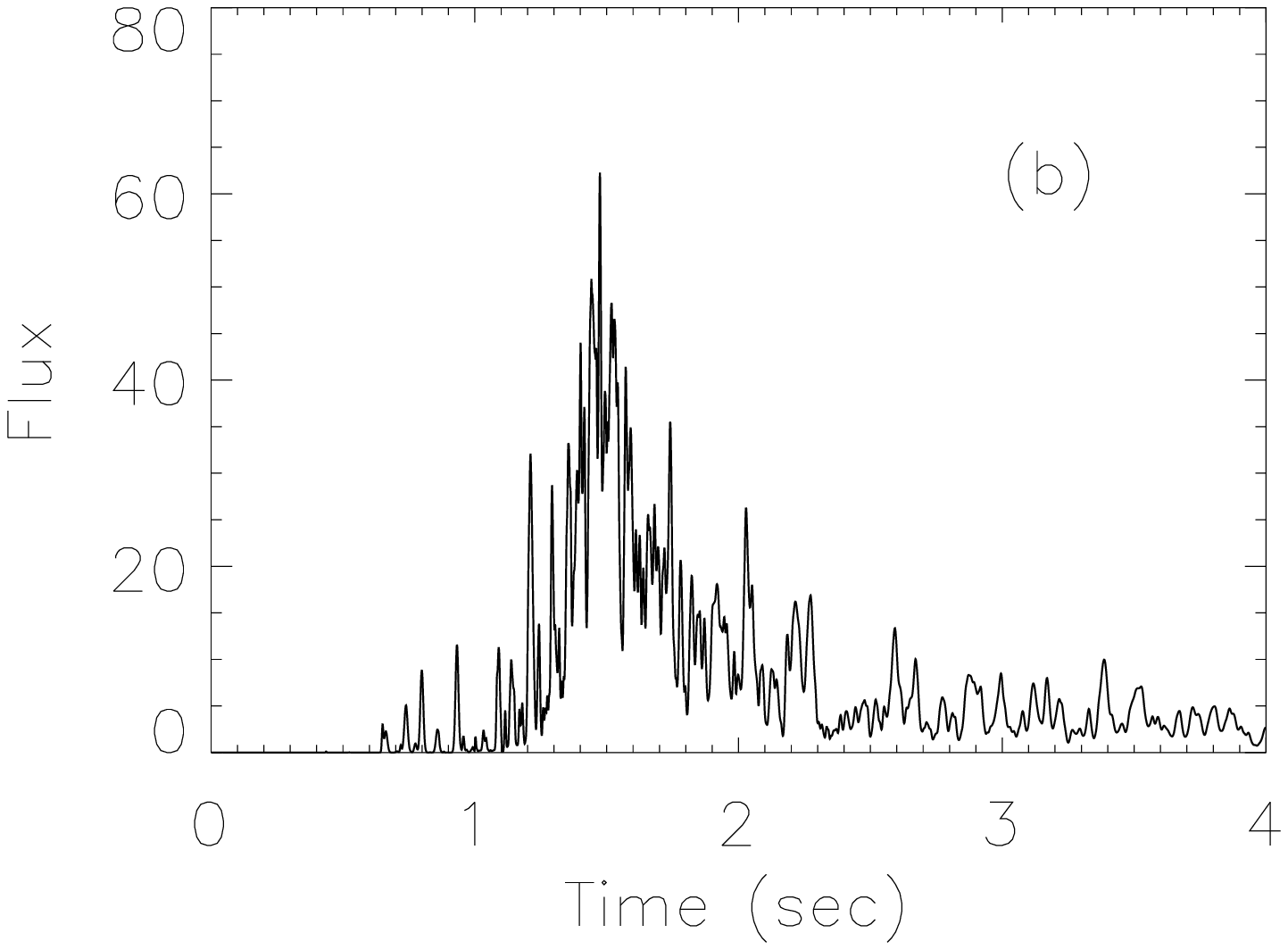}\\
\includegraphics*[width=7cm]{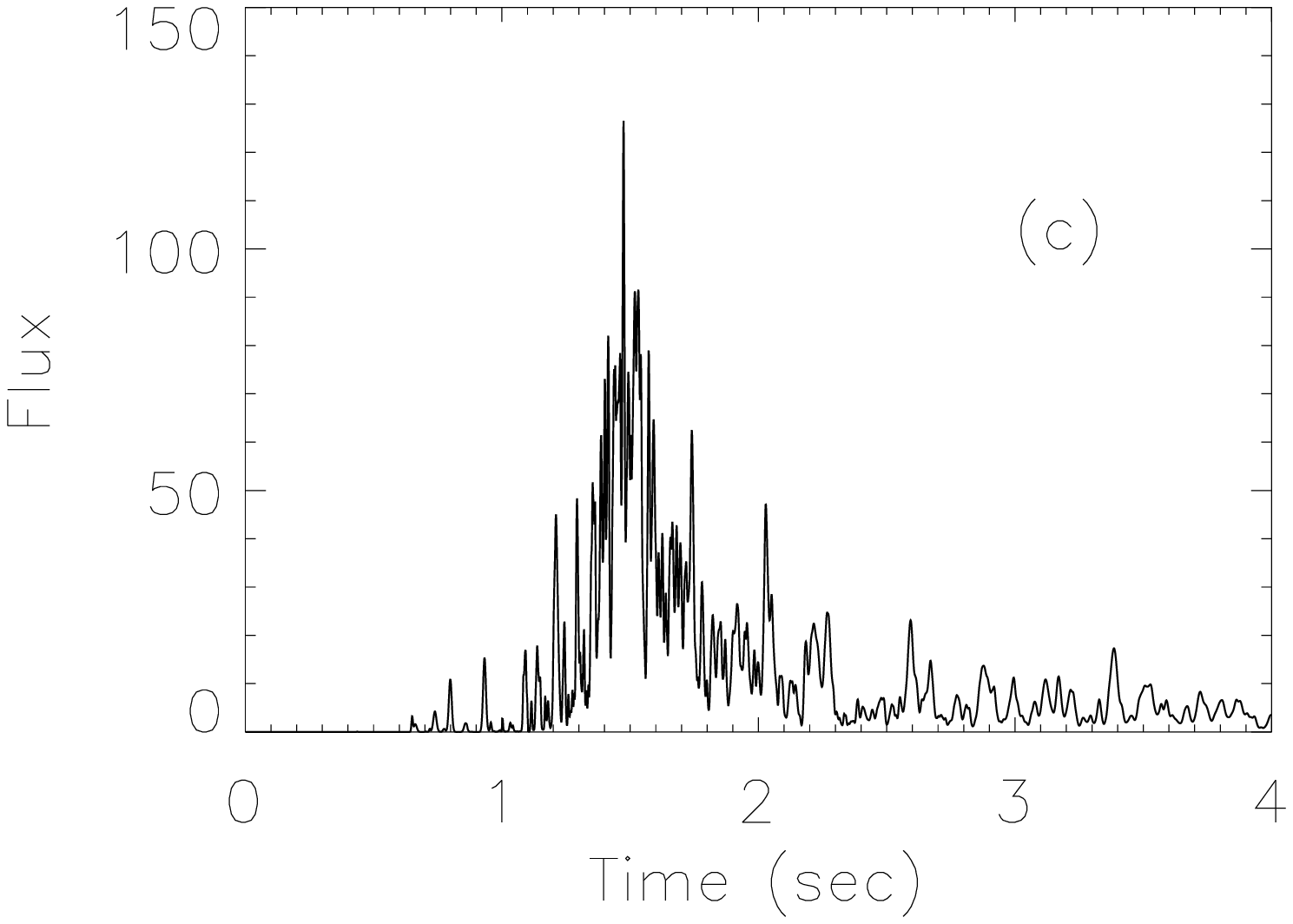}
\caption{\small{Simulated light curves of one ICMART event with
the following parameters: $R = 5\times 10^{15}$ cm,  $\Gamma_{\rm{ini}} = 200$,
$L^{'} = 5 \times 10^{11}$ cm, $N=50,000$ and rising time $t_r \sim 1.5$ s. 
The $\phi$-distribution is taken as Gaussian with typical angle $45^\circ$.
Three panels compare different $\gamma_{\rm{ini}}-\Gamma_{\rm{ini}}$ contrasts.
(a)  $\gamma_{\rm{ini}}=8$;
(b) $\gamma_{\rm{ini}}=14$; 
(c) $\gamma_{\rm{ini}} = 20$.}}\label{fig:gamma-ratio}
\end{center}
\end{figure}

\begin{figure}
\begin{center}
\includegraphics*[width=7cm]{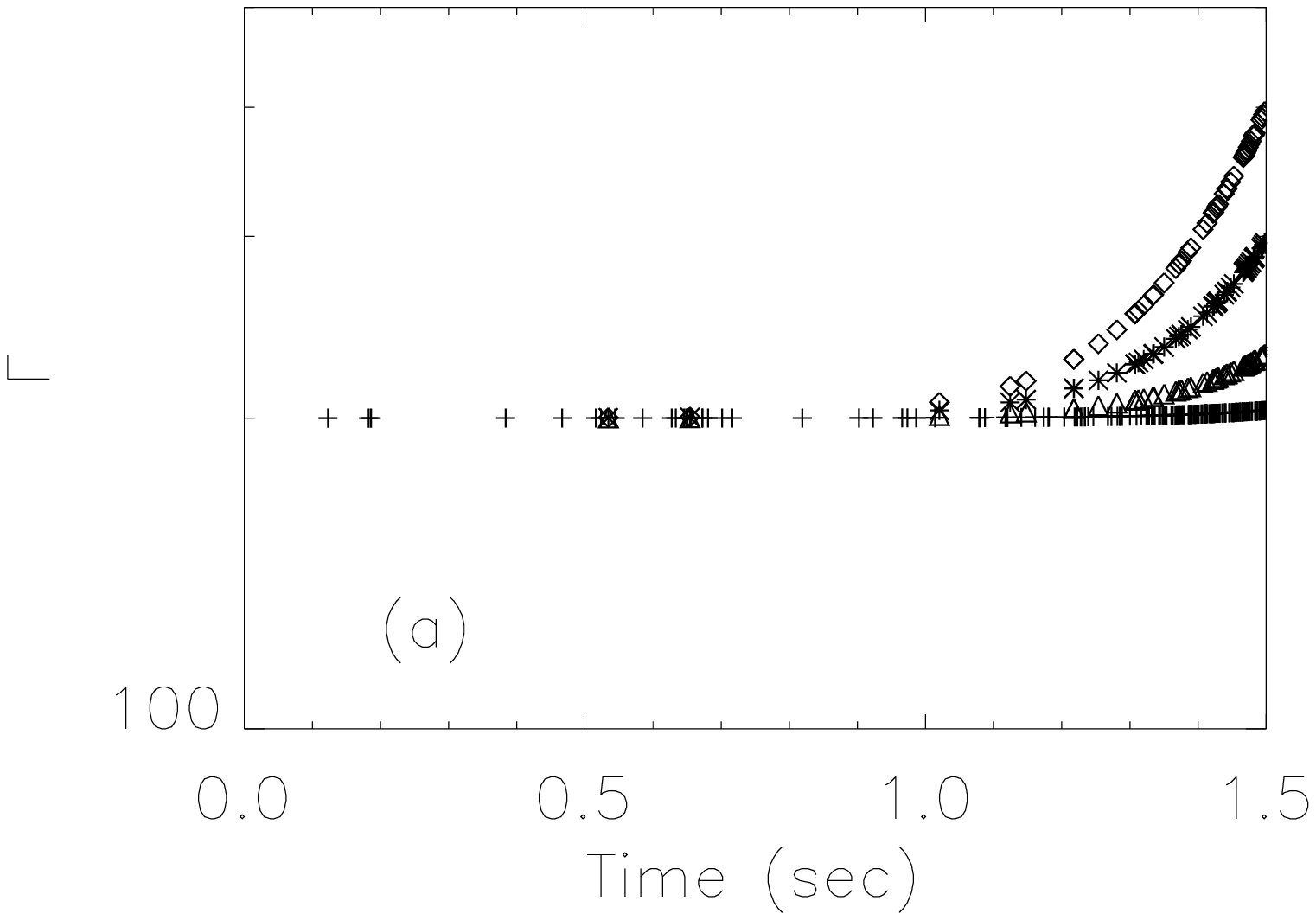}\\
\includegraphics*[width=7cm]{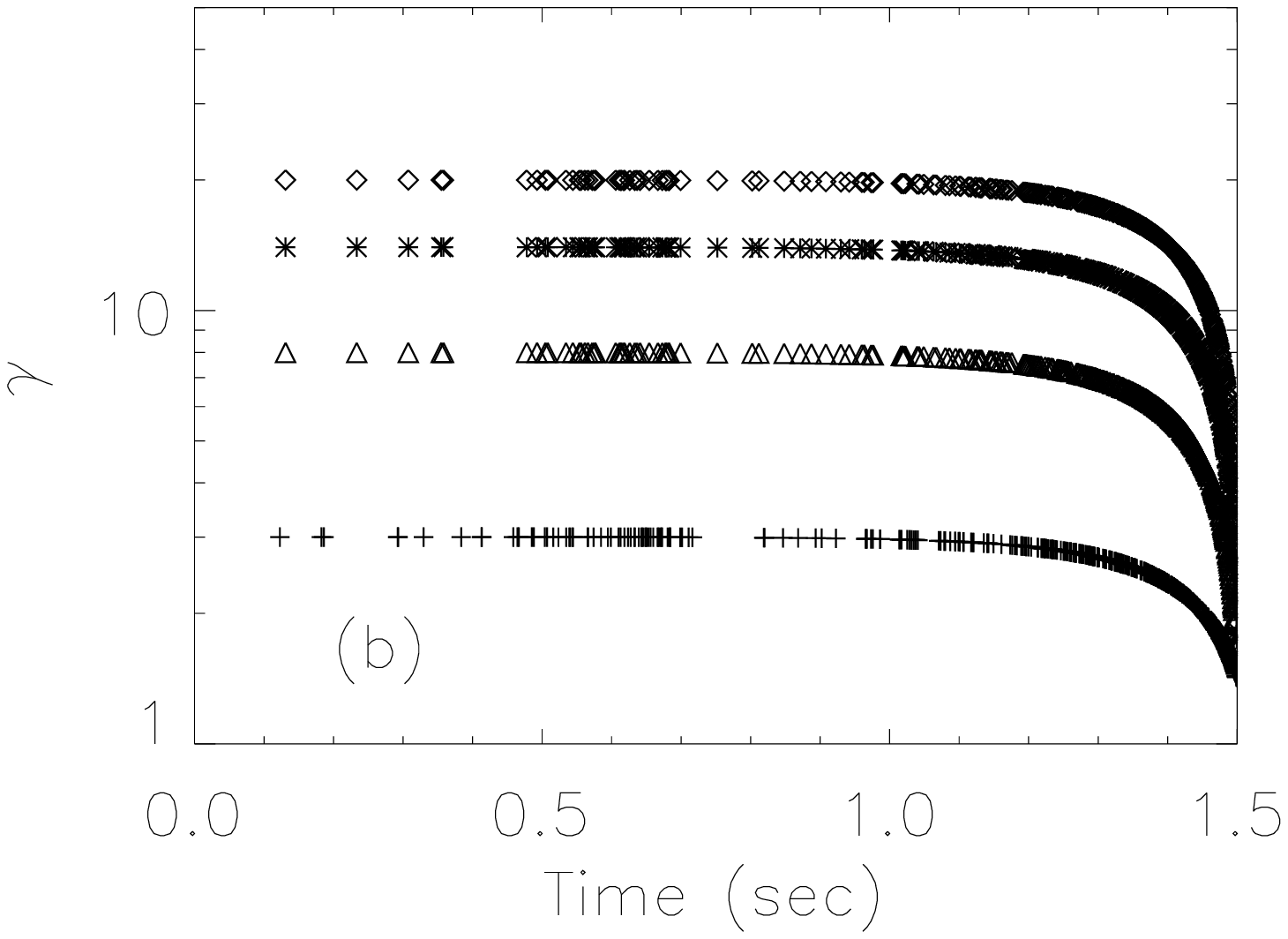}\\
\includegraphics*[width=7cm]{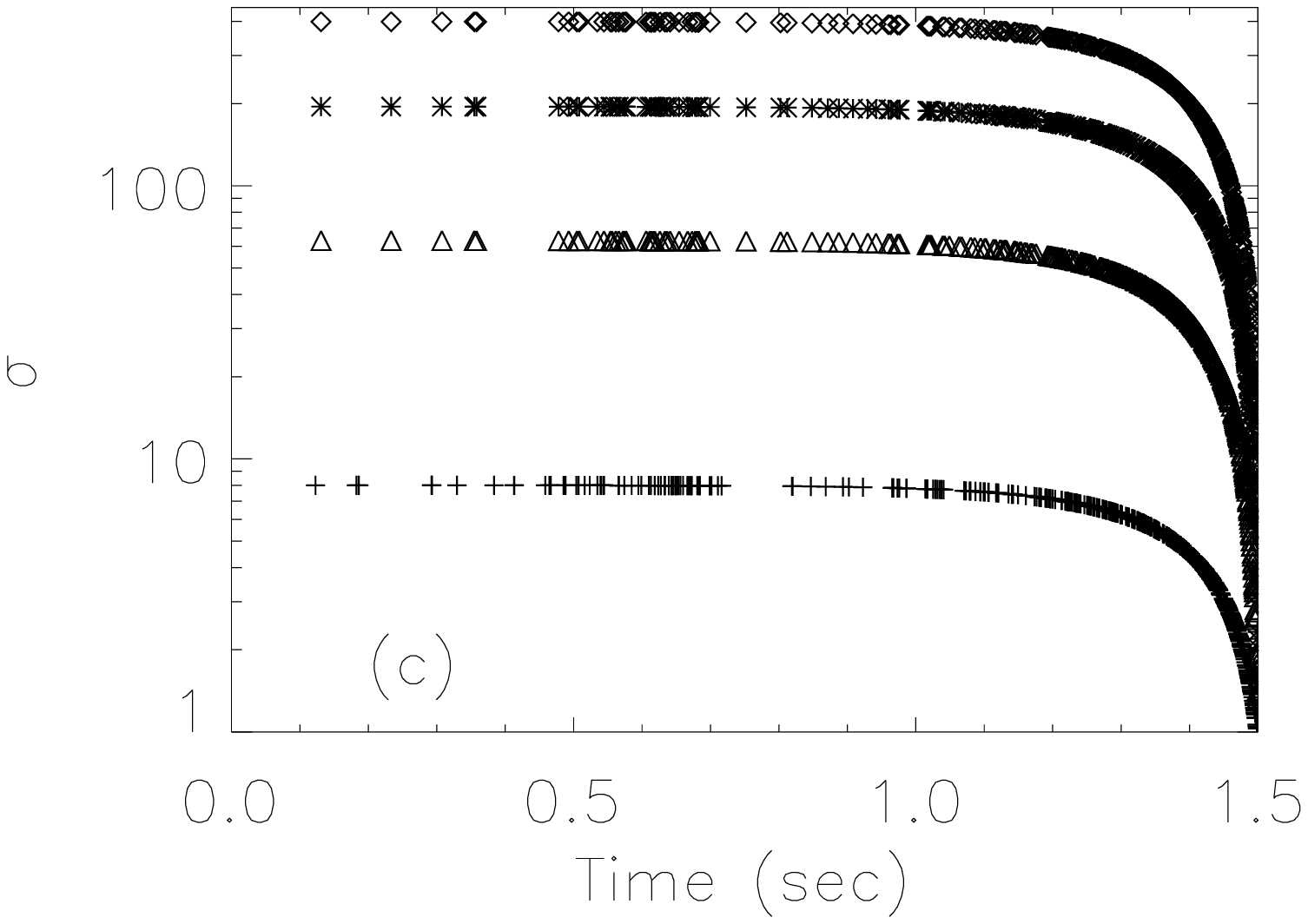}
\caption{\small{Evolution of the following parameters during the 
simulated ICMART event: (a) $\Gamma$; (b) $\gamma$; (c) $\sigma$.
The different symbols denote different parameters presented in Figures 1 and 2:
cross: $\Gamma_{\rm{ini}} = 200$, $\gamma_{\rm{ini}}=3$,  triangle: $\Gamma_{\rm{ini}} = 200$, $\gamma_{\rm{ini}}=8$, star: $\Gamma_{\rm{ini}} = 200$, $\gamma_{\rm{ini}}=14$, 
diamond: $\Gamma_{\rm{ini}} = 200$, $\gamma_{\rm{ini}}=20$.}}\label{fig:evolution}
\end{center}
\end{figure}

We next compare the effect of Lorentz factor contrast in the ICMART 
region. We keep the initial value of the bulk Lorentz factor
$\Gamma$ constant, i.e.,
$\Gamma_{\rm{ini}} = 200$, and vary $\gamma_{\rm{ini}}$. 
This corresponds to different values of the 
initial magnetization $\sigma_{\rm{ini}}$. In Figure \ref{fig:gamma-ratio}, we 
compare three sets of simulations, with (a) $\gamma_{\rm{ini}}=8$; 
(b) $\gamma_{\rm{ini}} = 14$ and
(c) $\gamma_{\rm{ini}}=20$.
Other parameters are the same as those adopted to calculate
Figure \ref{fig:direction}, and the Gaussian $\phi$-distribution model with
typical angle $45^\circ$ has been adopted.
We show that the light curves become progressively more erratic
and spikier when the $\gamma_{\rm{ini}}$  becomes larger. This is 
because a larger $\gamma_{\rm{ini}}$ would give rise to larger ${\cal D}_2$,
and thus a larger value 
of the total Doppler factor ${\cal D}_1{\cal D}_2$. 
A larger $\gamma_{\rm{ini}}$ also tends to give a more significant evolution of 
the parameters (Figure \ref{fig:evolution}). Initially, a constant $\Gamma_{\rm{ini}}$ corresponds to 
a constant $1/\Gamma_{\rm{ini}}$ cone, so that observed numbers of mini-jets are
the same in all these cases. However a larger $\gamma_{\rm{ini}}$ can give rise to
a larger $\Gamma$ near the end of evolution, thus a smaller 
$1/\Gamma$ cone. 
The slow component is not as significant, so that the light curves
become spikier.

In order to show the evolution of the physical parameters during
the ICMART cascade event, in Figure \ref{fig:evolution} we display 
the evolution of the bulk Lorentz 
factor $\Gamma$, the mini-jet Lorentz factor $\gamma$, and the
emission region magnetization $\sigma$ as a function of time.
It can be seen that evolution is more significant for a larger
$\gamma_{\rm{ini}}$ (and equivalently a larger $\sigma_{\rm{ini}}$). 

\subsubsection{Number of Reconnection Events}

Next, we test how the total number of mini-jets $N$ within the $1/\Gamma$ cone
affects the light curves.  According to Equation \ref{eq:N}, varying $N$ is effectively
varying the filling factor $f$. By varying $N$, the total number of e-folding 
steps $n$ is slightly modified, as is the rising time $t_r$. 
In Figure 4, we compare the simulated light curves for different $N$ values,
i.e., $N = 10^4, 5 \times 10^4, 10^5$, and $5\times 10^5$, respectively. 
It can be seen that in general the light curves appear smoother with 
increasing $N$. 
This can be readily understood: the larger the $N$, the more reconnection events
happen simultaneously, so that more mini-jets beaming to different directions 
tend to enhance the slow component. The short-timescale
structures are smeared out, and the light curves become smoother.

\begin{figure}
\begin{center}
\includegraphics*[width=4cm]{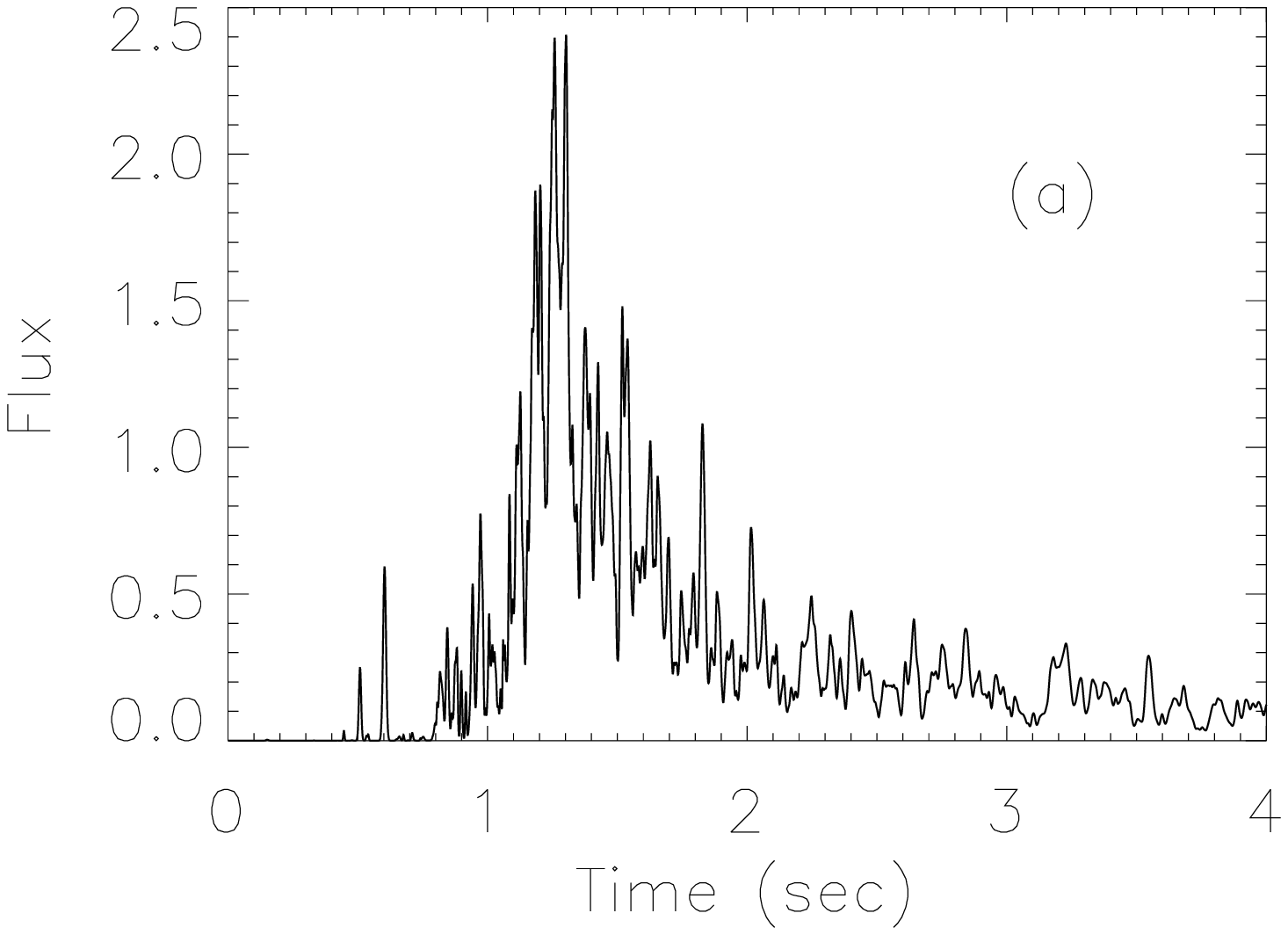}
\includegraphics*[width=4cm]{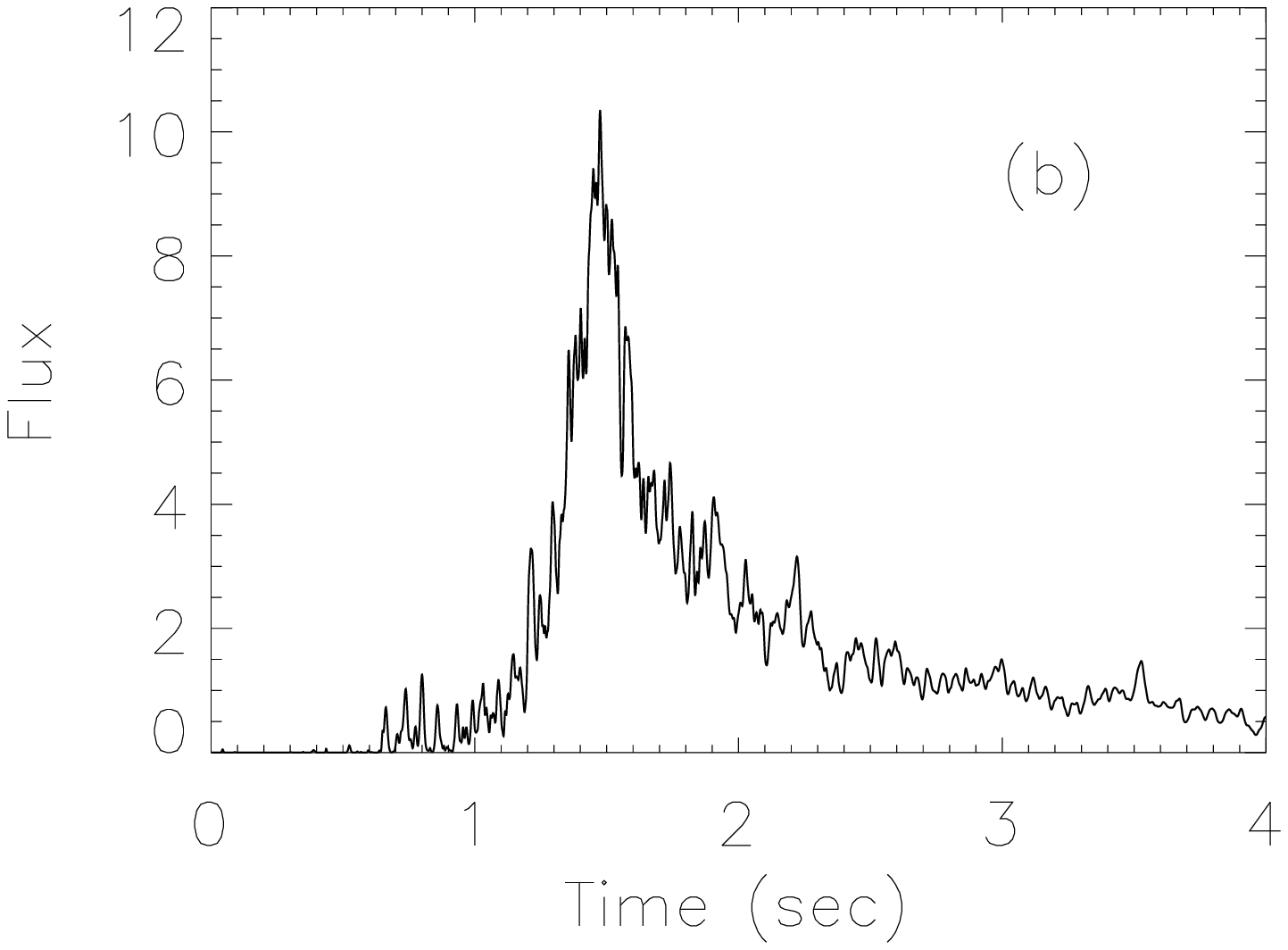}
\\
\includegraphics*[width=4cm]{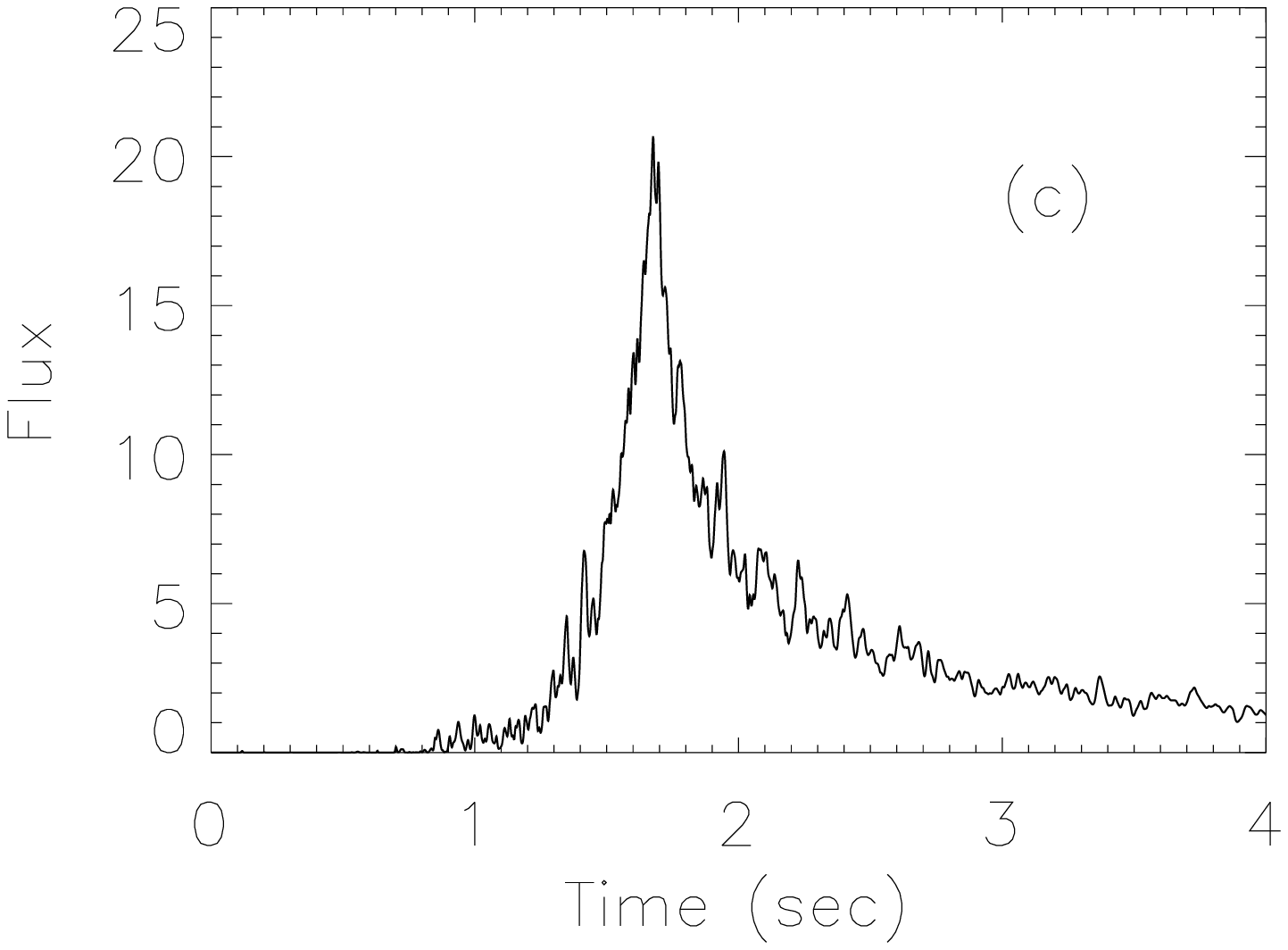}
\includegraphics*[width=4cm]{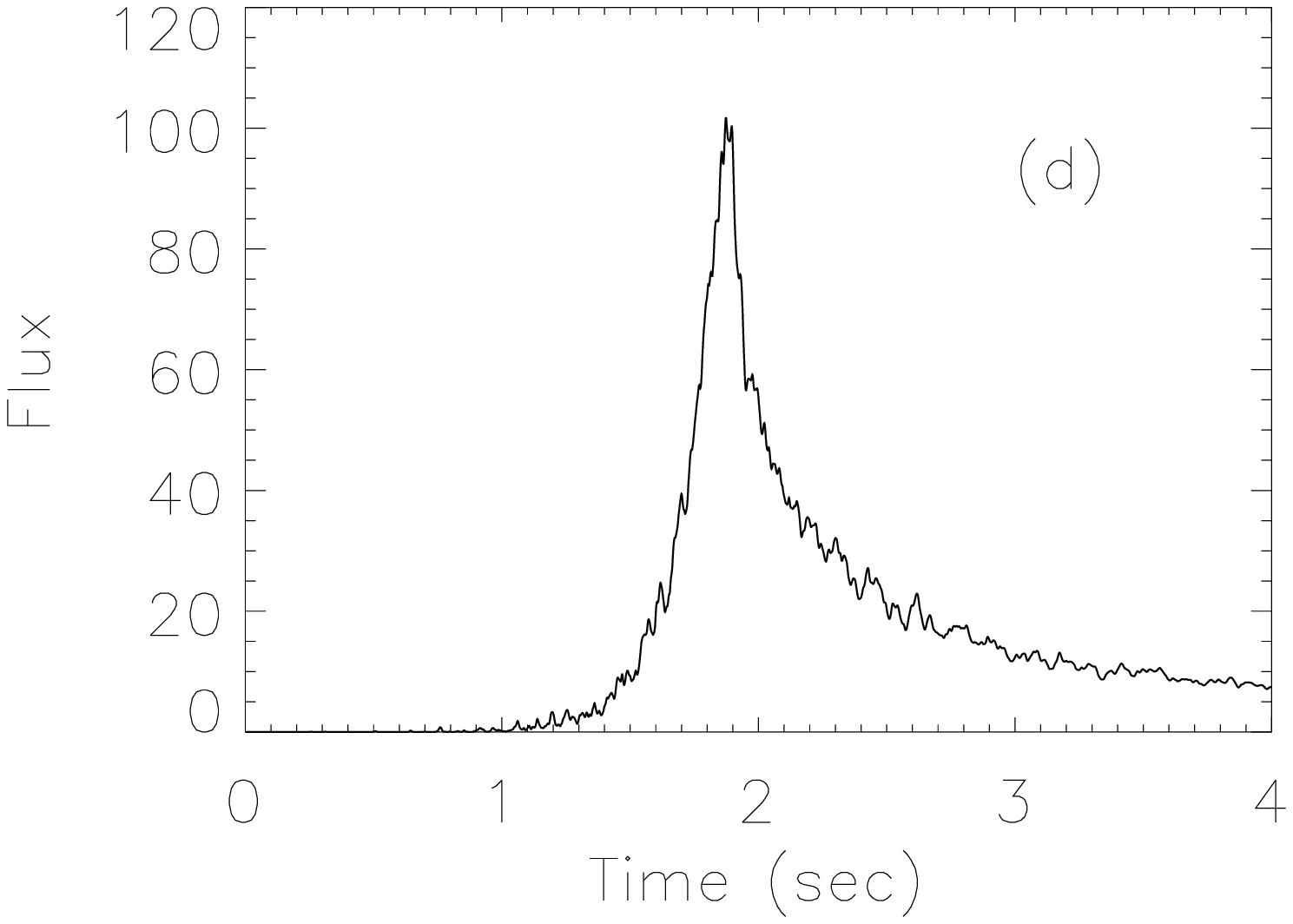}
\caption{\small{A comparison of the cases with different reconnection events $N$
The parameters are the same as those 
in Figure 1 except $N$. 
(a) $N = 10^4$; (b) $N = 5\times 10^4$;
(c) $N = 10^5$; (d) $N = 5\times 10^5$.}}\label{fig:N}
\end{center}
\end{figure}

\subsubsection{Emission Radius}\label{sec:R}

\begin{figure}
\begin{center}
\includegraphics*[width=7cm]{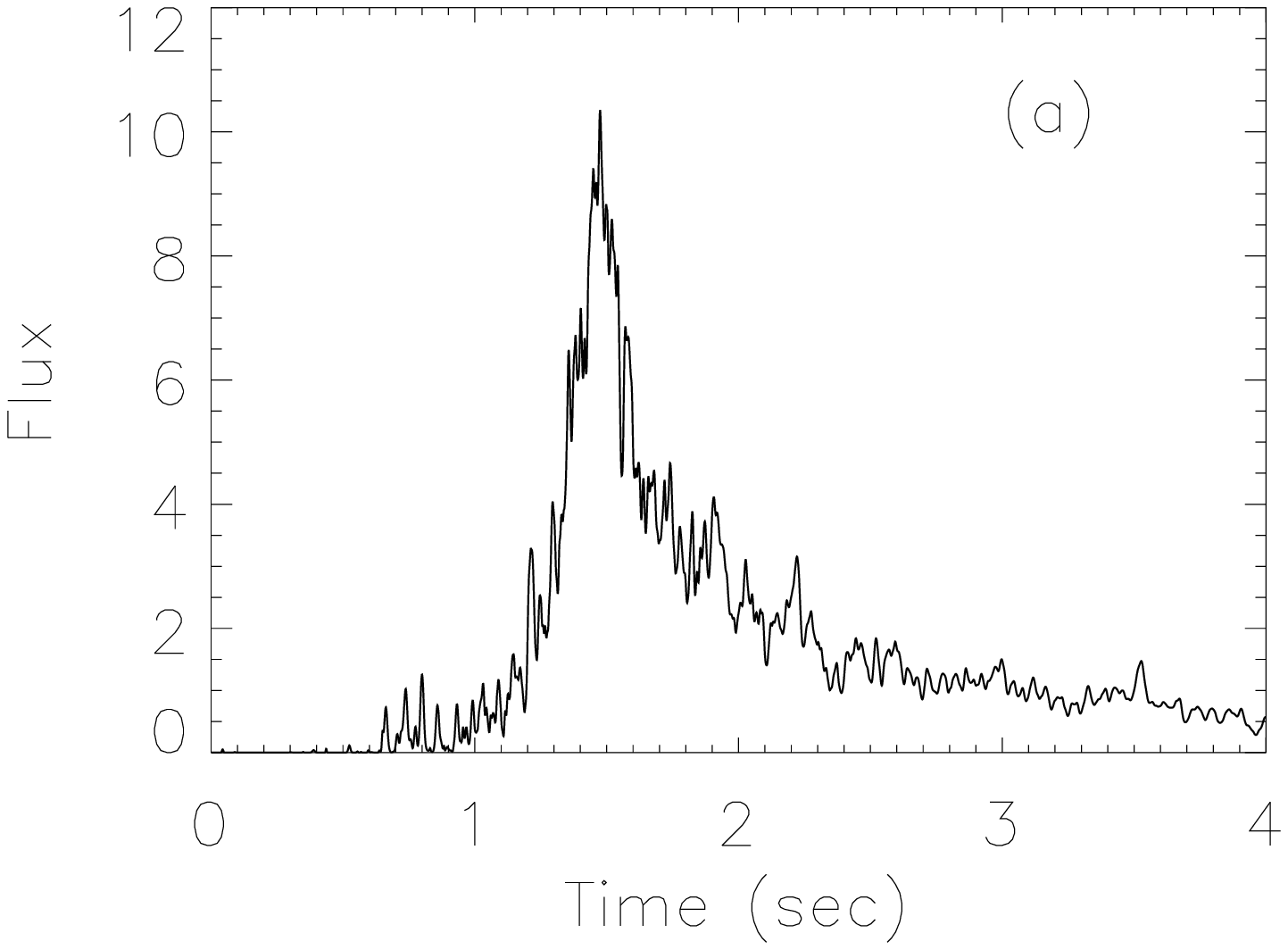}\\
\includegraphics*[width=7cm]{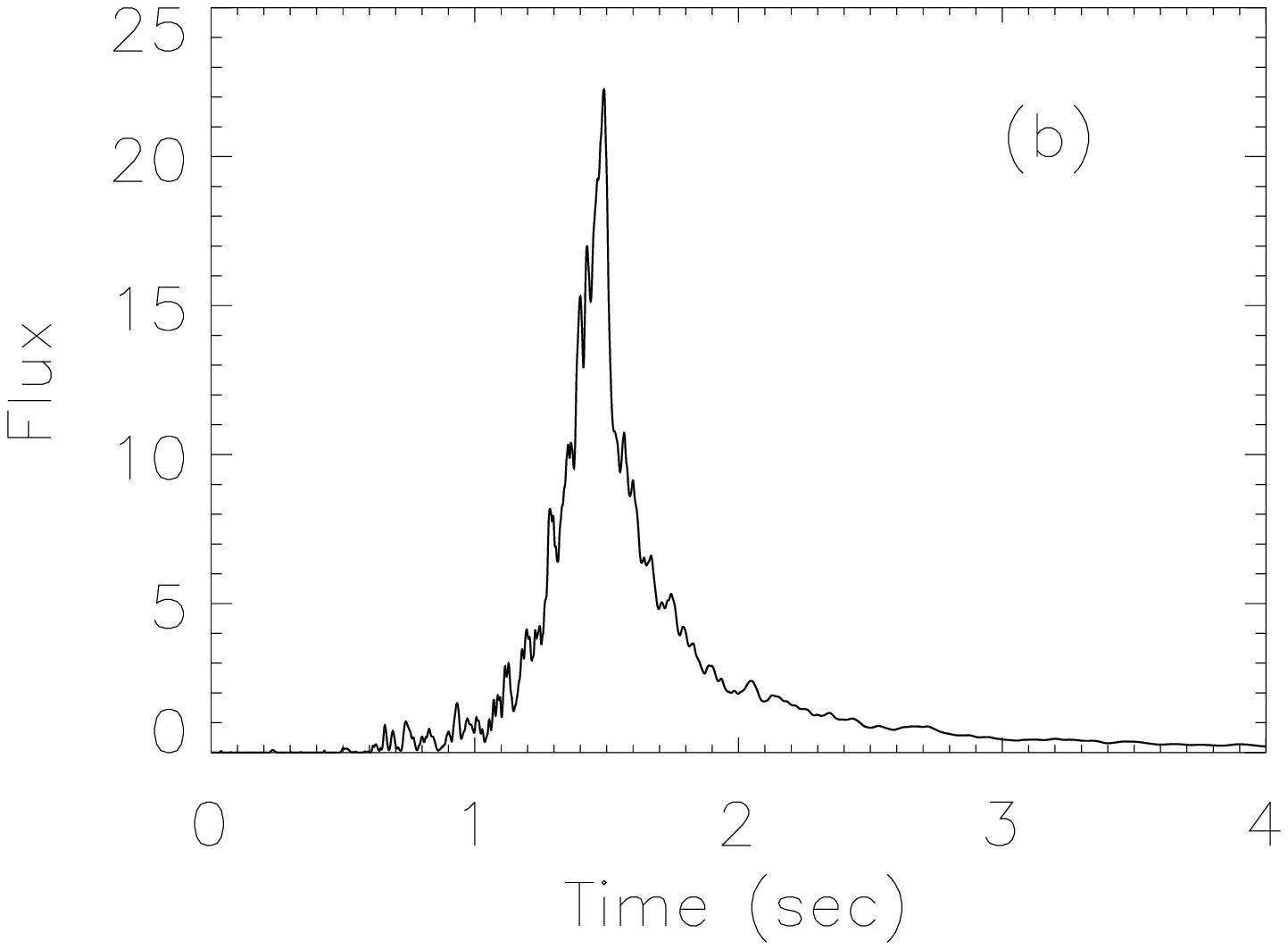}\\
\includegraphics*[width=7cm]{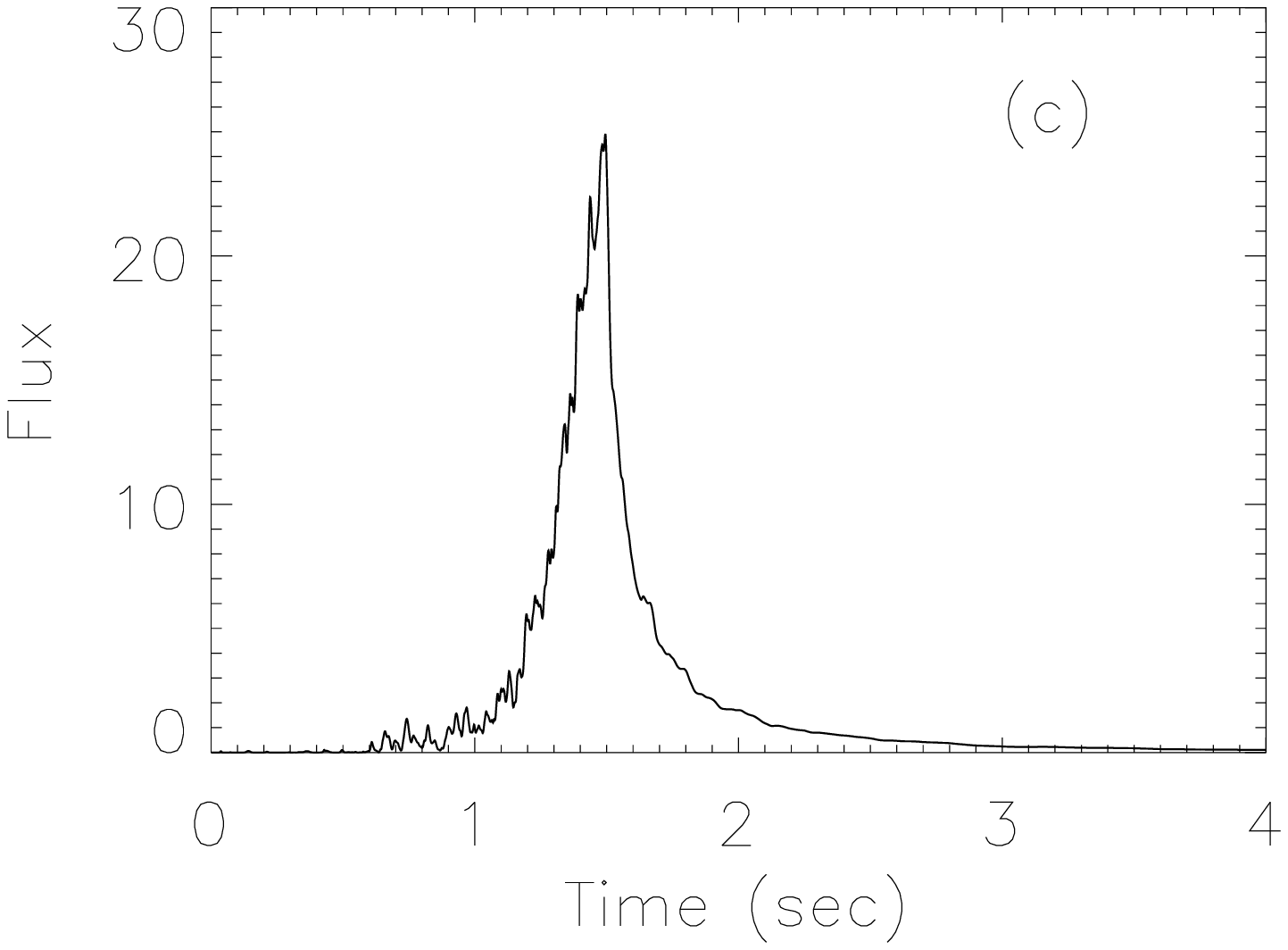}
\caption{\small{The simulated light curves with different emission 
region radius $R$: (a) $R = 5\times 10^{15}$ cm, (b) $R = 10^{15}$ cm,
and (c) $R = 5\times 10^{14}$ cm. The other parameters are the same as those 
in Figure 1.}}\label{fig:R}
\end{center}
\end{figure}

Next, we explore the effect of the emission region radius $R$. 
Figures \ref{fig:R}(a), \ref{fig:R}(b), and \ref{fig:R}(c) 
show the results for $R = 5\times 10^{15}$ cm, 
$R = 10^{15}$ cm, and $R = 5\times 10^{14}$, respectively. One can see 
that the larger the $R$, the longer and stronger the high-latitude emission tail.
This is because the length of the high-latitude tail is defined
by $R(1-\cos\theta_j)$. We notice again that the rising time is
the growth time of the cascade, which is the $e$-folding time of 
consuming most of the magnetic energy in the emission region, which
is defined by the total number $N$ of the mini-jets and the 
characteristic scale $L'$ of each mini-jet. Since the rising and
falling times are related to different parameters, the pulse is usually
asymmetric (e.g., Figure \ref{fig:R}). The simulated light curve is more
consistent with data if the emission radius $R$ is large. 
ZY11 suggested that ICMART events should happen at larger 
radii, say, $R > 10^{15}$ cm, in order to reach the critical condition
of triggering a reconnection/turbulence cascade. It is intriguing to
see that such large-radius ICMART events make light curves 
more resemble the observed ones.

\subsubsection{Size of the Reconnection Regions}

\begin{figure}
\begin{center}
\includegraphics*[width=7cm]{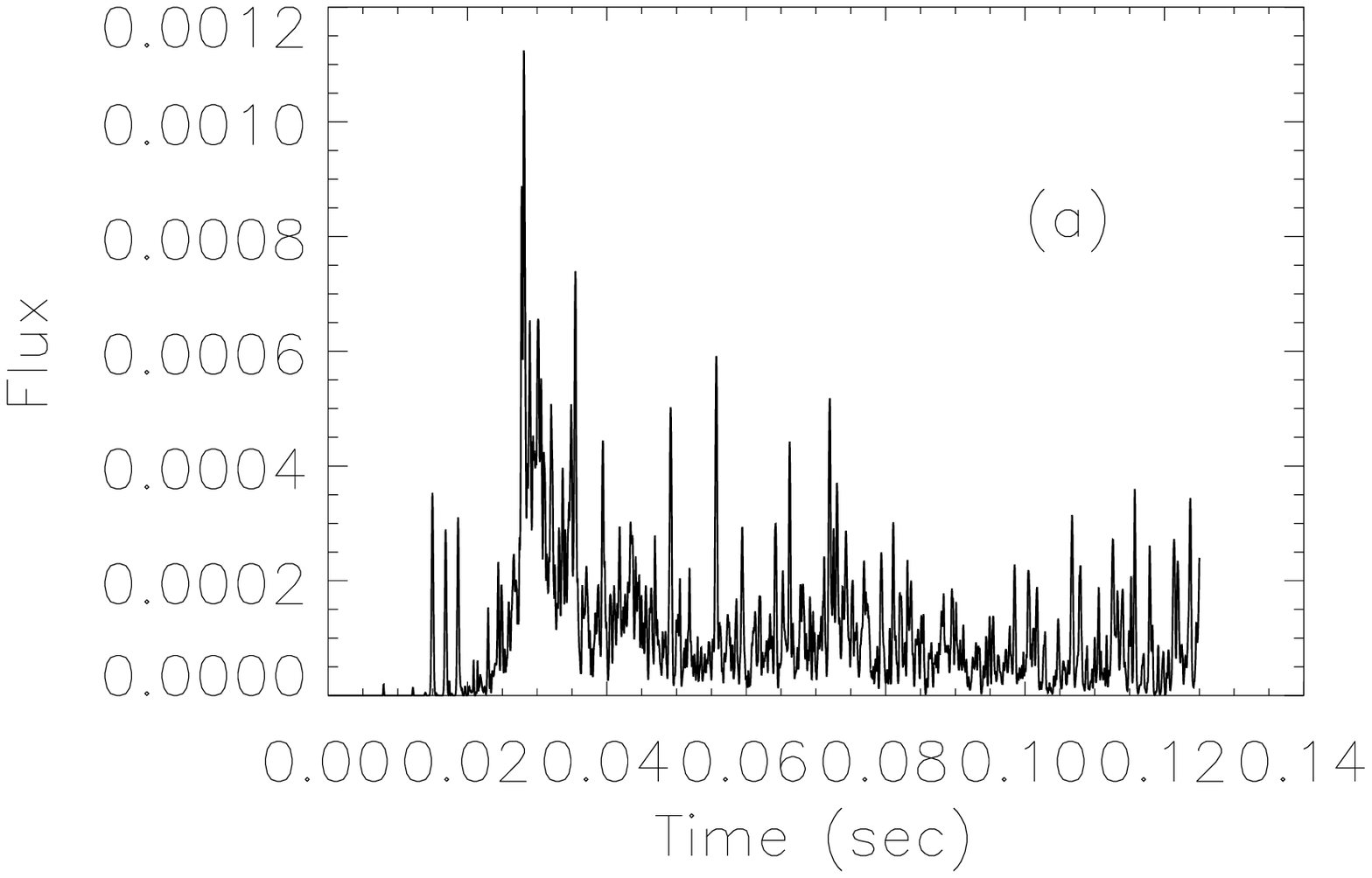}\\
\includegraphics*[width=7cm]{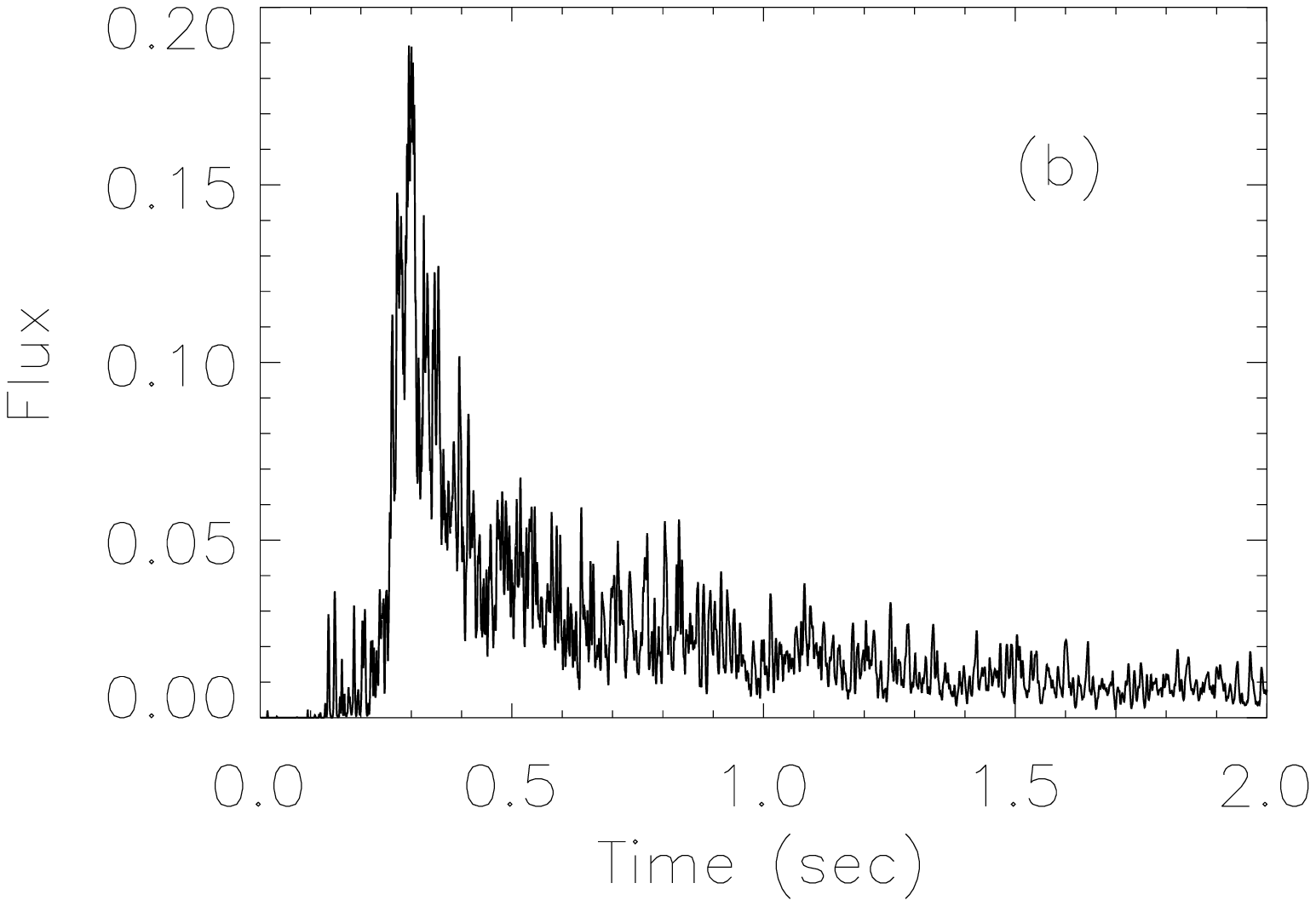}\\
\includegraphics*[width=7cm]{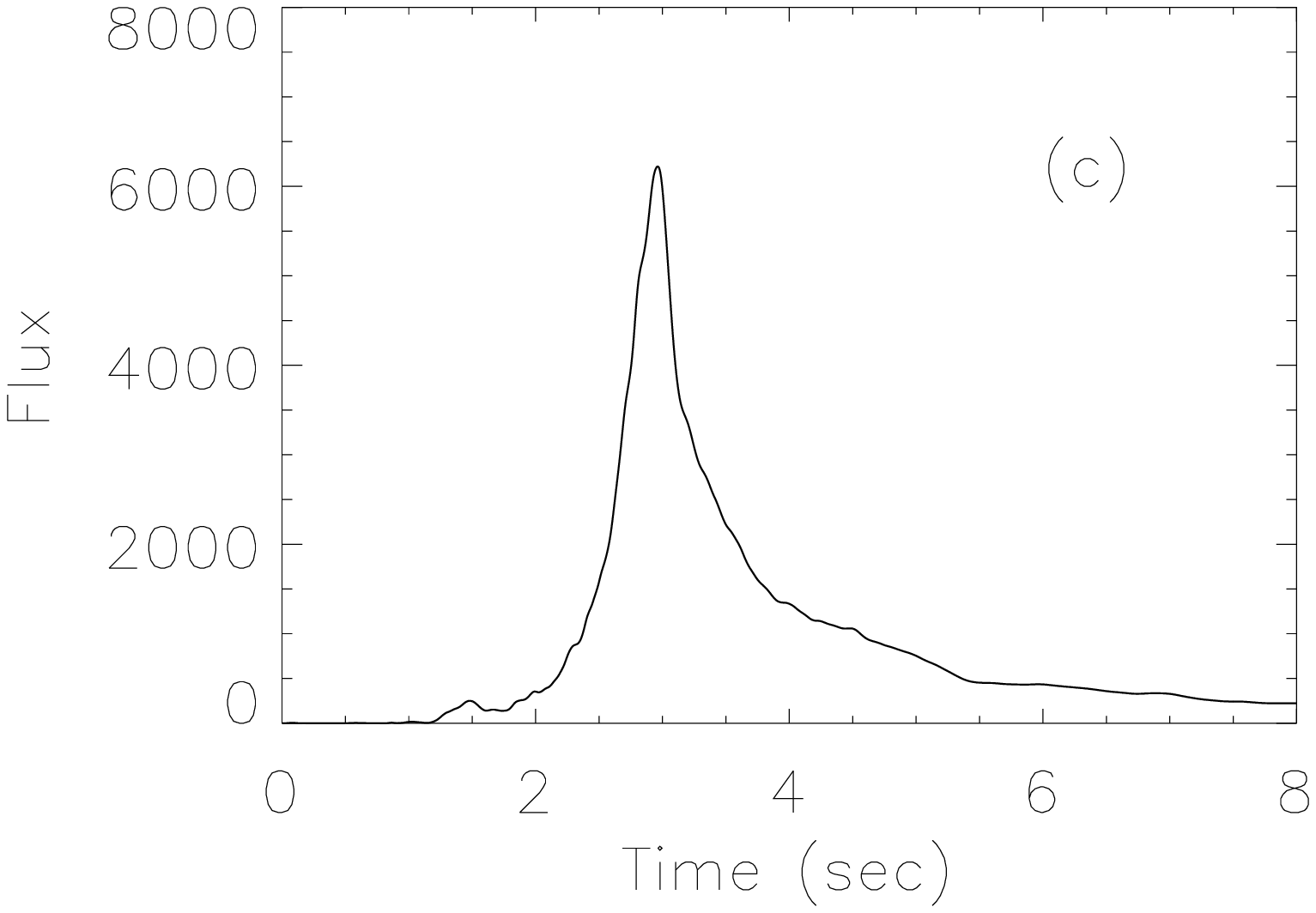}
\caption{\small{The simulated light curves with different size of reconnection region  $L'$: (a) $L' =  10^{10}$ cm, (b) $L' = 10^{11}$ cm,
and (c) $L' =  10^{12}$ cm. The other parameters are the same as those 
in Figure 1.}}\label{fig:size1}
\end{center}
\end{figure}

We also discuss the effect of different sizes of reconnection regions. We make
two sets of simulations. In the first set, we vary $L'$ while keeping $R$ constant.
We also keep $N = 50,000$, so effectively, we are varying the filling factor $f$.
Since $t_0 = L'/v_{\rm{in}}$, the rising time $t_{\rm{r}}$ is modified correspondingly. 
The results are presented in Figure \ref{fig:size1}, which shows the simulated 
light curves for $L' = 10^{10}$, $10^{11}$, and $10^{12}$ cm, respectively. 
It can be seen that the smaller the $L'$, the spikier the light curve. This
is because a smaller $L'$ corresponds to a shorter duration of each
reconnection event. For the $L' = 10^{12}$ cm case, short-timescale structures 
are missing, and the light curve is very smooth.

\begin{figure}
\begin{center}
\includegraphics*[width=7cm]{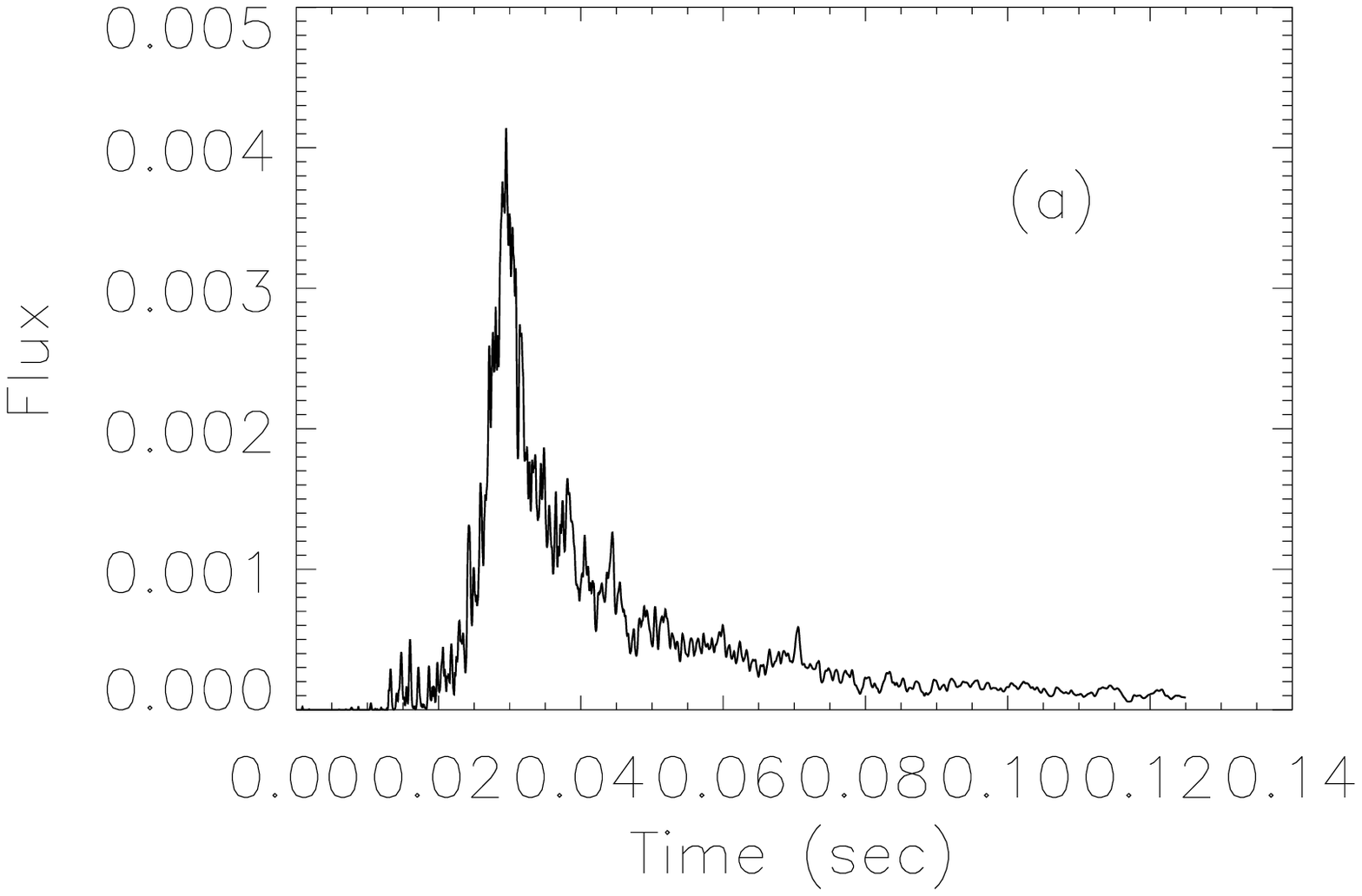}\\
\includegraphics*[width=7cm]{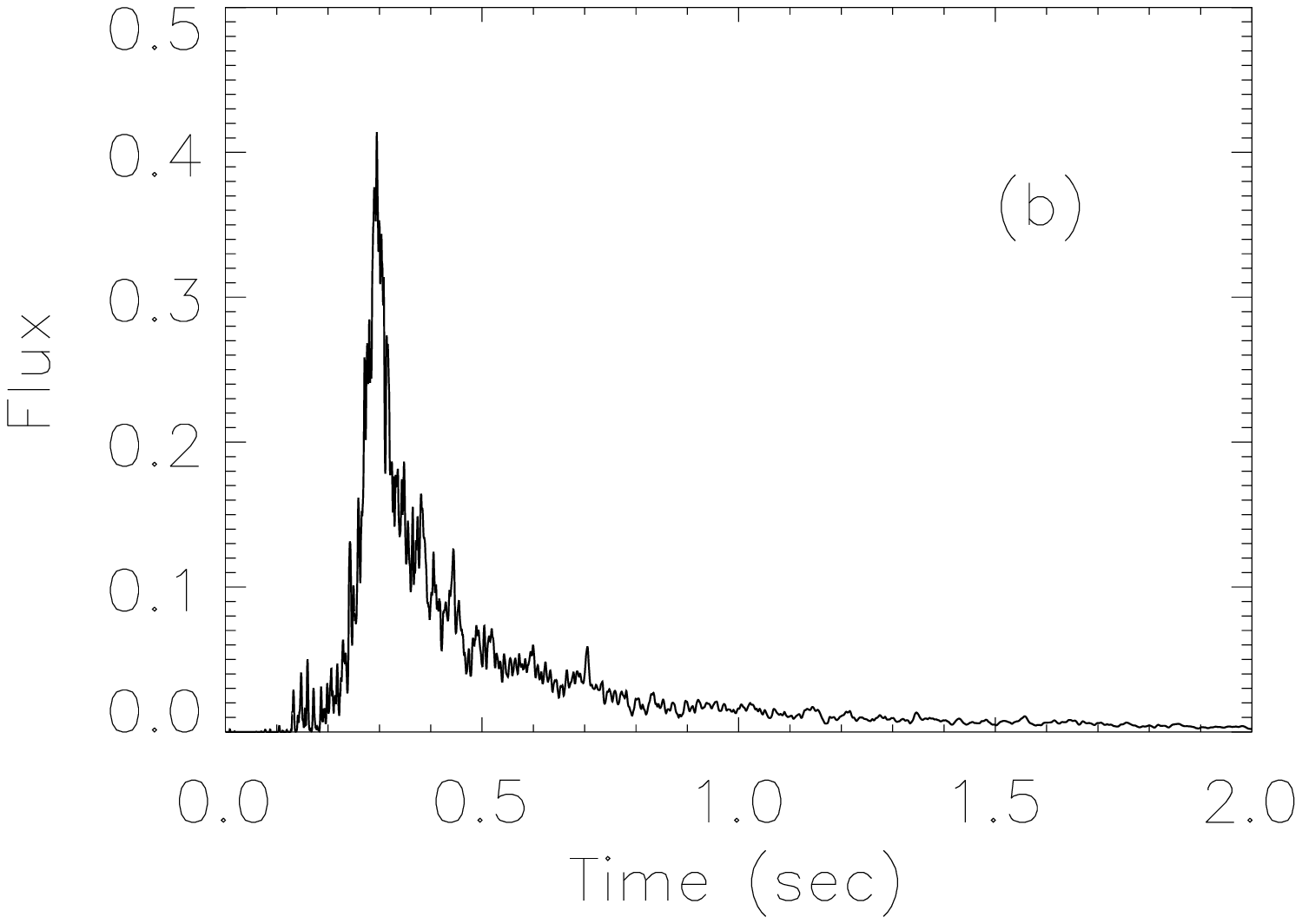}\\
\includegraphics*[width=7cm]{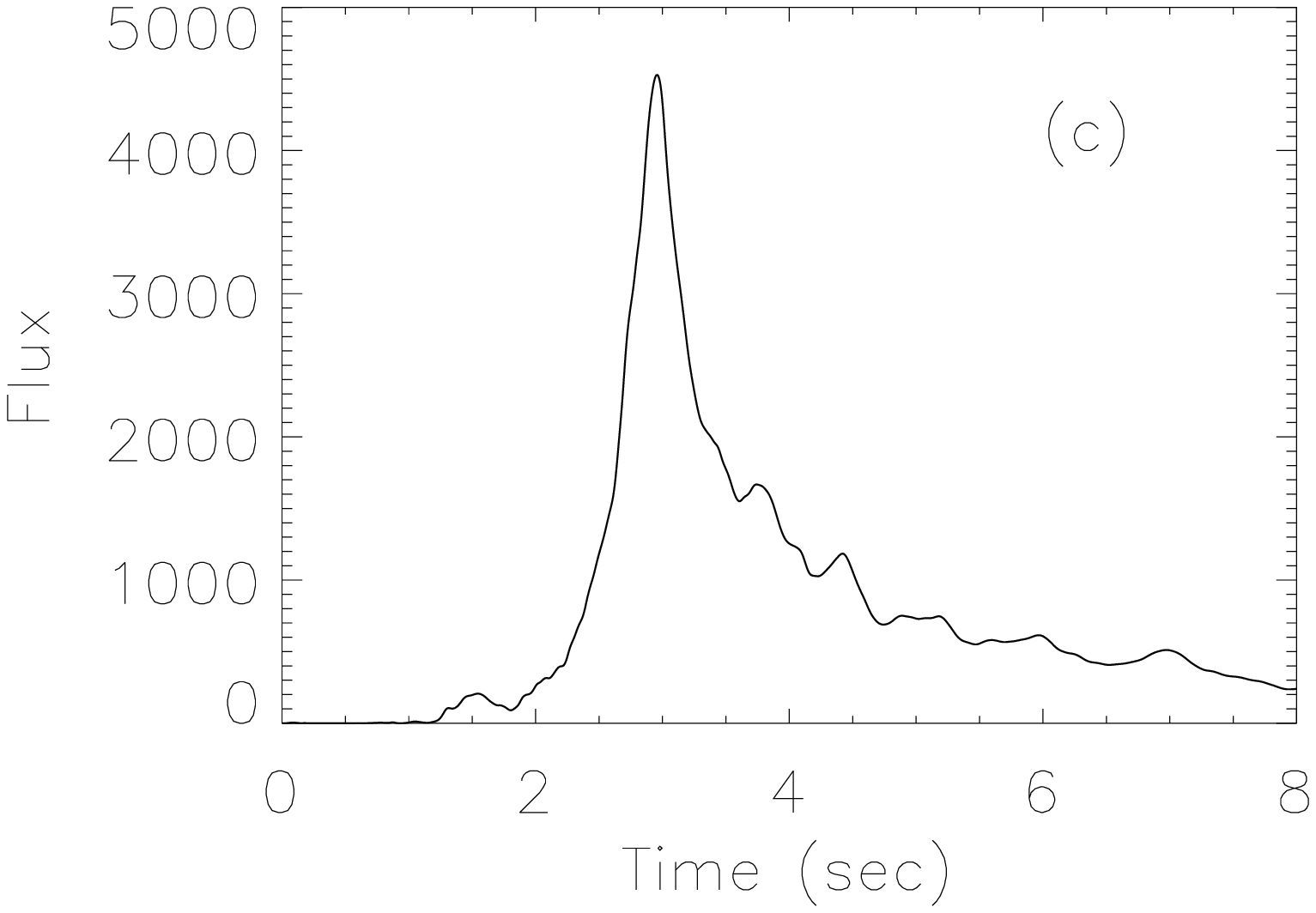}
\caption{\small{The simulated light curves with different size of reconnection region $L'$, while keeping $L'/R$ constant: (a) $L' =  10^{10}$ cm, (b) $L' = 10^{11}$ cm,
and (c) $L' =  10^{12}$ cm. The other parameters are the same as those 
in Figure 1.}}\label{fig:size2}
\end{center}
\end{figure}

Next, we keep both $N$ and $f$ constant. By varying $L'$, we are effectively
varying $R$ as well, so that the ratio $L'/R$ is a constant. The results are
shown in Figure \ref{fig:size2}, in which light curves for  $L' = 10^{10}$, 
$10^{11}$, and $10^{12}$ cm are simulated. The general trend as discussed 
above is still there, but since $R$ is changed accordingly, the contrasts are
less significant, namely, the smaller $L'$ cases are less spiky and larger
$L'$ cases are less smooth with respect to the case where $R$ is fixed
(Figure \ref{fig:size1}. Since the decay phase is defined by $R$ (Section \ref{sec:R}
above), varying $R$ with $L'$ also affects the length of the decaying phase.

\subsubsection{Size distribution}

\begin{figure}
\begin{center}
\includegraphics*[width=7cm]{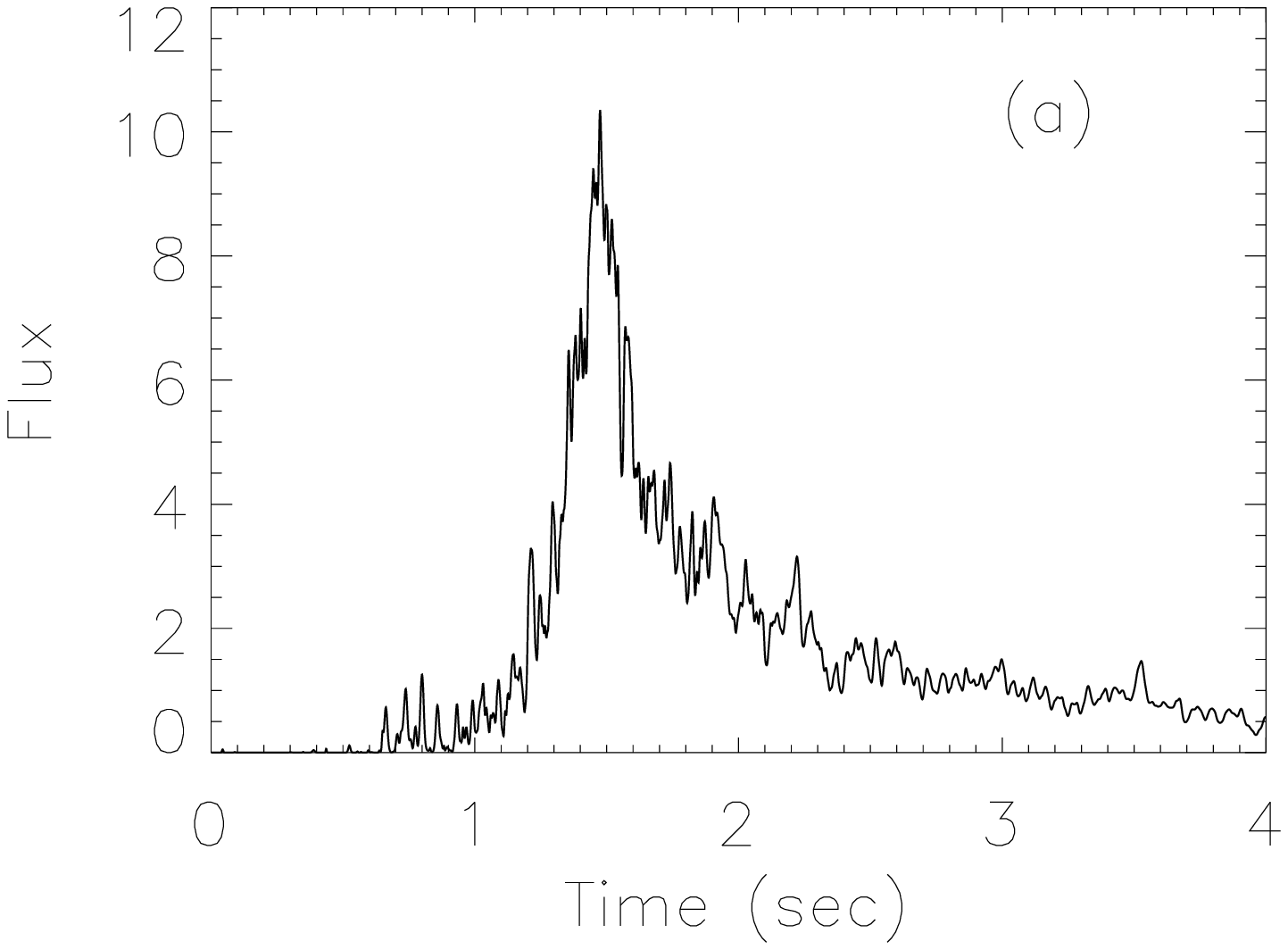}\\
\includegraphics*[width=7cm]{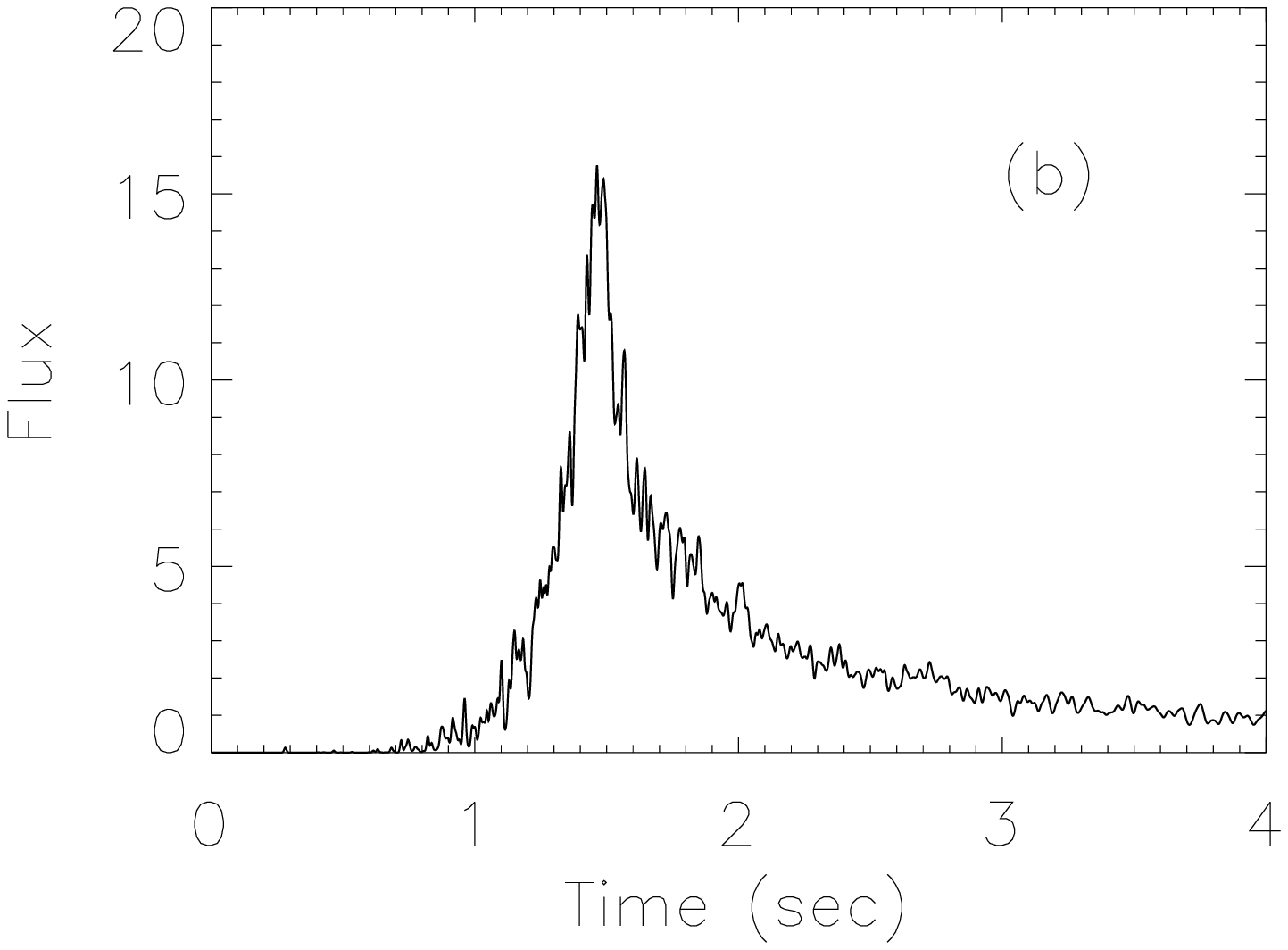}
\caption{\small{The simulated light curves with different reconnection region size 
distribution: (a) power law distribution with an index of  $-5/3$
; (b) uniform distribution. The other parameters are the same as those in Fig.1.}}
\label{fig:size-distribution}
\end{center}
\end{figure}

We next test the effect of size distribution of the reconnection 
regions. We try two possibilities: the power-law distribution with
an index $-5/3$ (the Kolmogorov type) (Figure \ref{fig:size-distribution}(a)) and a uniform 
distribution (Figure \ref{fig:size-distribution}(b)). One can see that the uniform distribution has 
a smoother shape. In this case, the observed small pulse width 
distribution is solely determined by the distribution of the Doppler
factors. For the power-law distribution case, an extra factor (the
intrinsic distribution) plays a role to make small pulses, so that
the light curves are spikier. 

\subsubsection{Energy dependence}

\begin{figure}
\begin{center}
\includegraphics*[width=7cm]{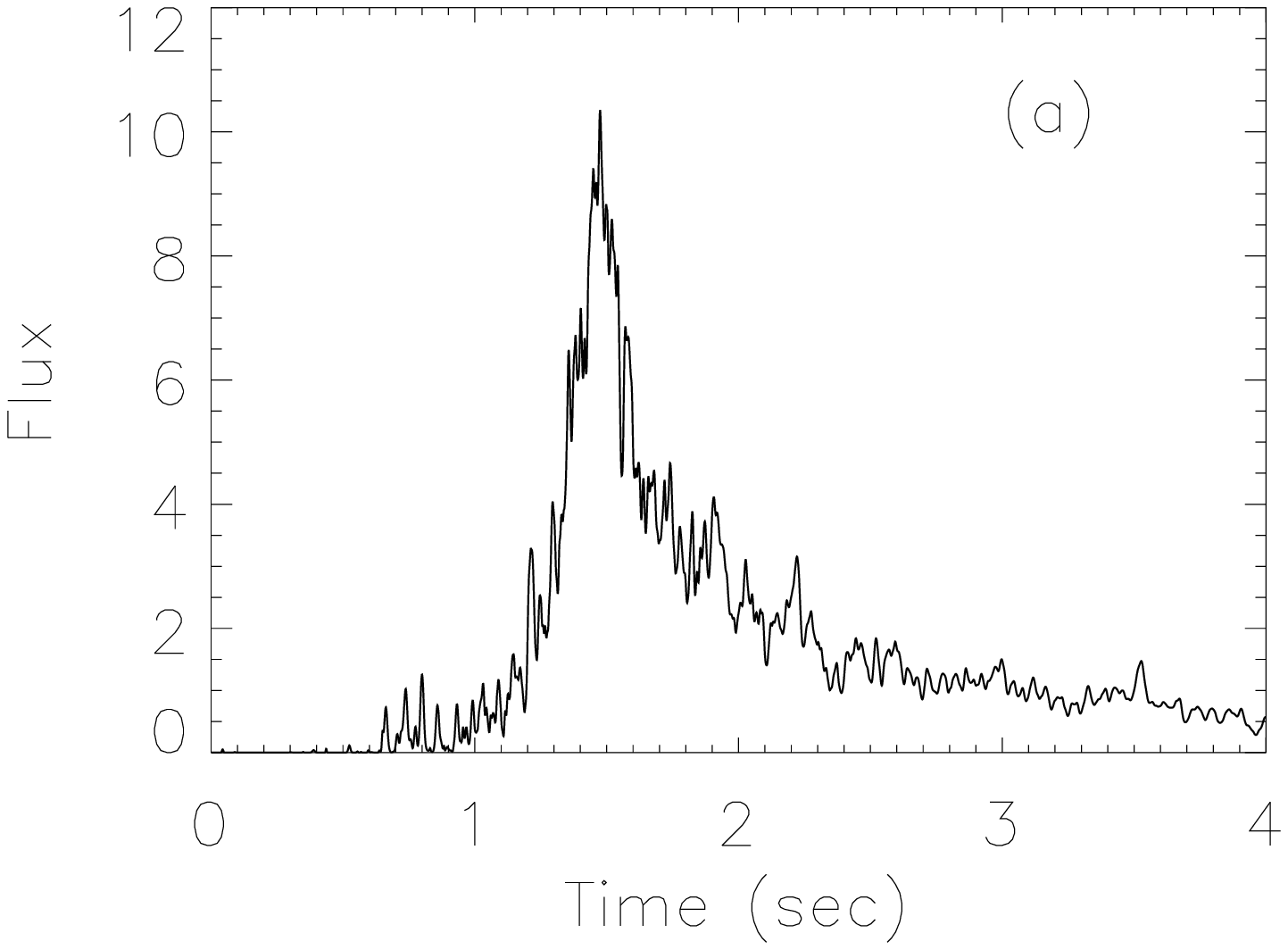}\\
\includegraphics*[width=7cm]{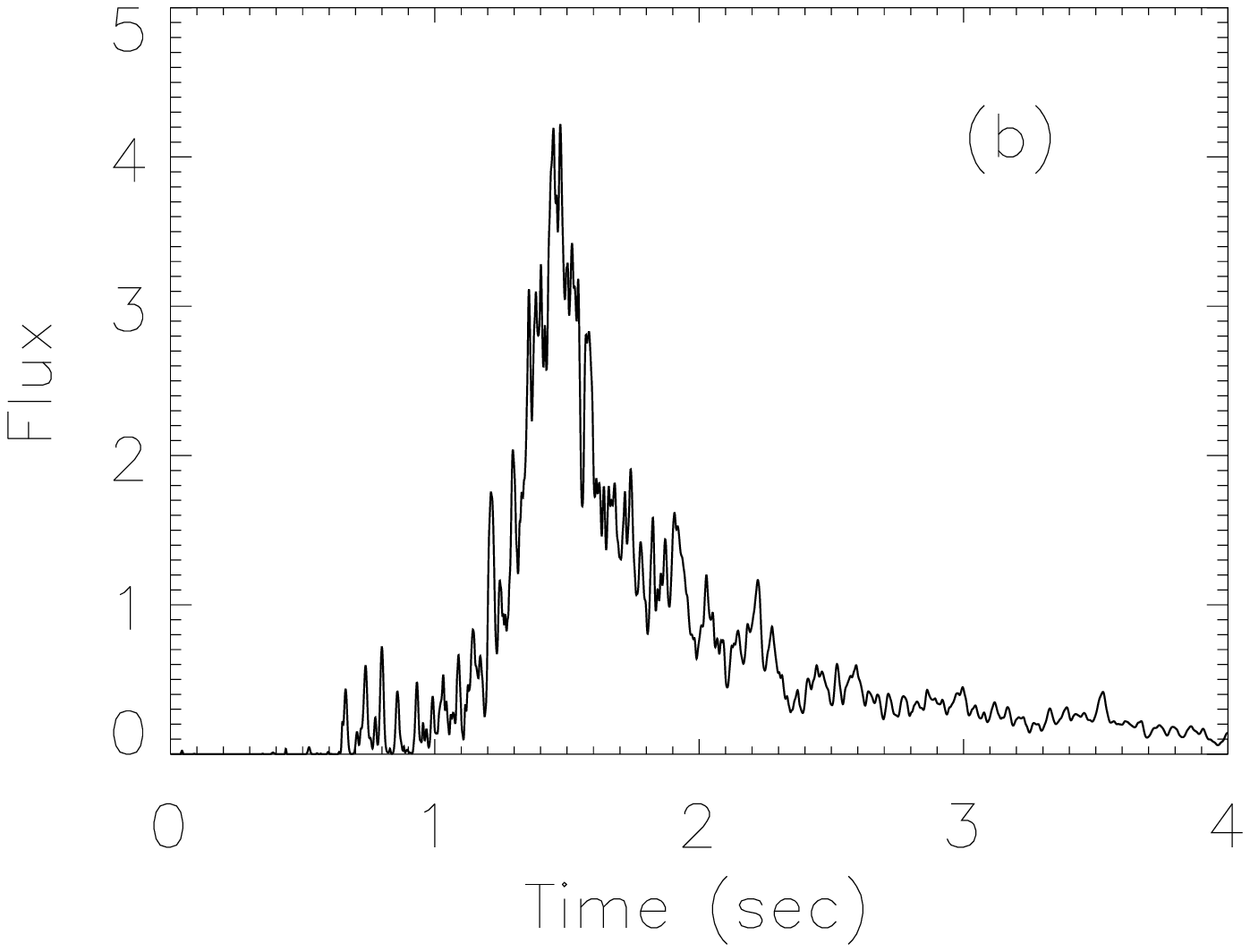}\\
\includegraphics*[width=7cm]{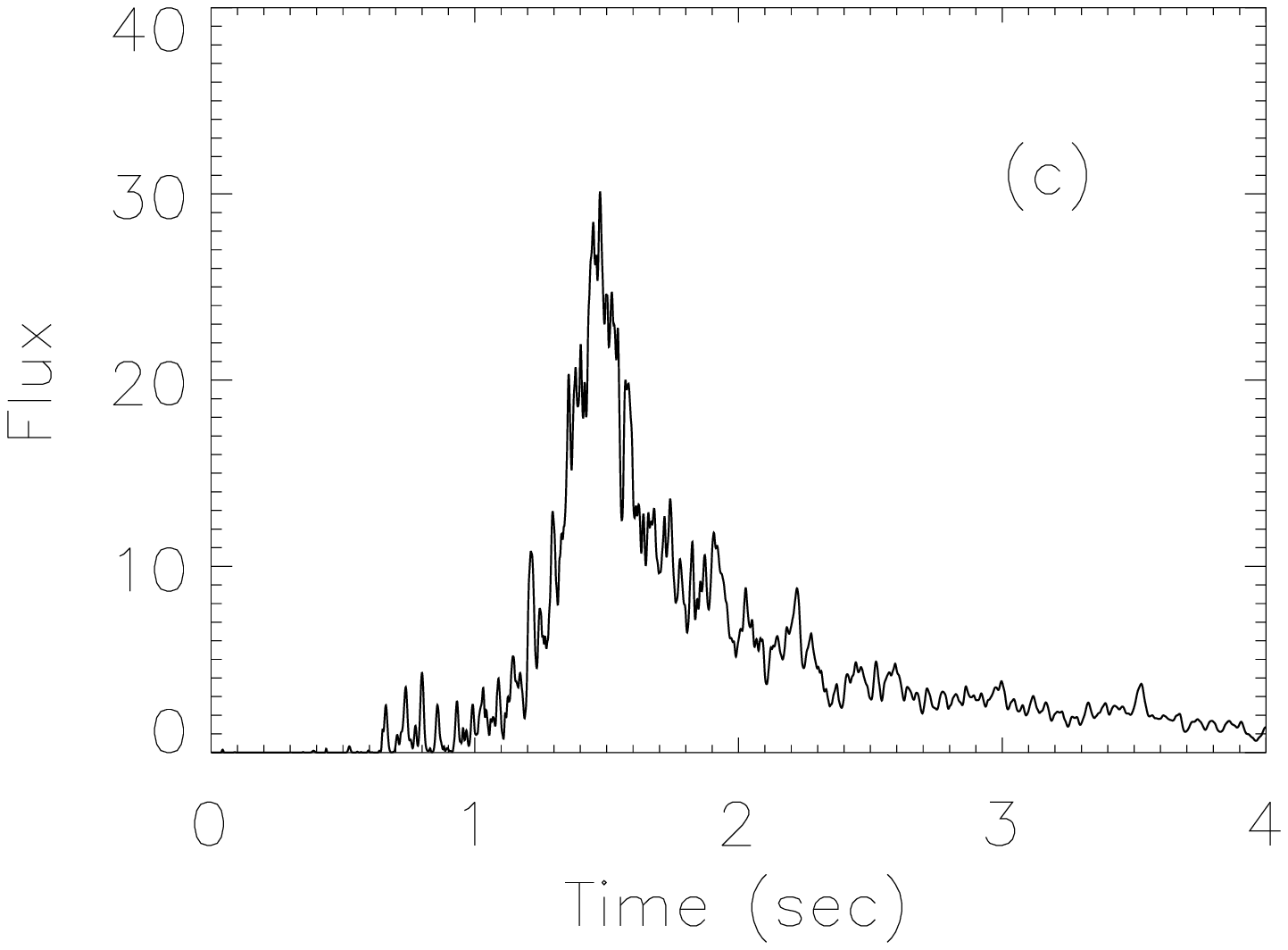}
\caption{\small{The simulated light curves in different observing bands: 
(a) 15 - 150 keV; (b) 500 - 650 keV; (c) 15 - 650 keV. The other parameters 
are the same as those in Figure 1.}}\label{fig:bands}
\end{center}
\end{figure}

Finally we calculate the light curves for different energy bands. We 
consider three cases here, below the peak of the band spectrum (Figure \ref{fig:bands}(a)),
i.e., 15 - 150 keV (also the observation band for \textit{Swift} BAT), 
above the peak (500 - 650 keV, Figure \ref{fig:bands}(b)), and across the entire energy band 
(15 - 650 keV, Figure \ref{fig:bands}(c)). The high-energy light curve is slightly narrower
and spikier, as observed in real GRBs. In general, the overall shape of 
the light curves does not differ significantly.

\subsubsection{Multiple episodes}

\begin{figure}
\begin{center}
\includegraphics*[width=8cm]{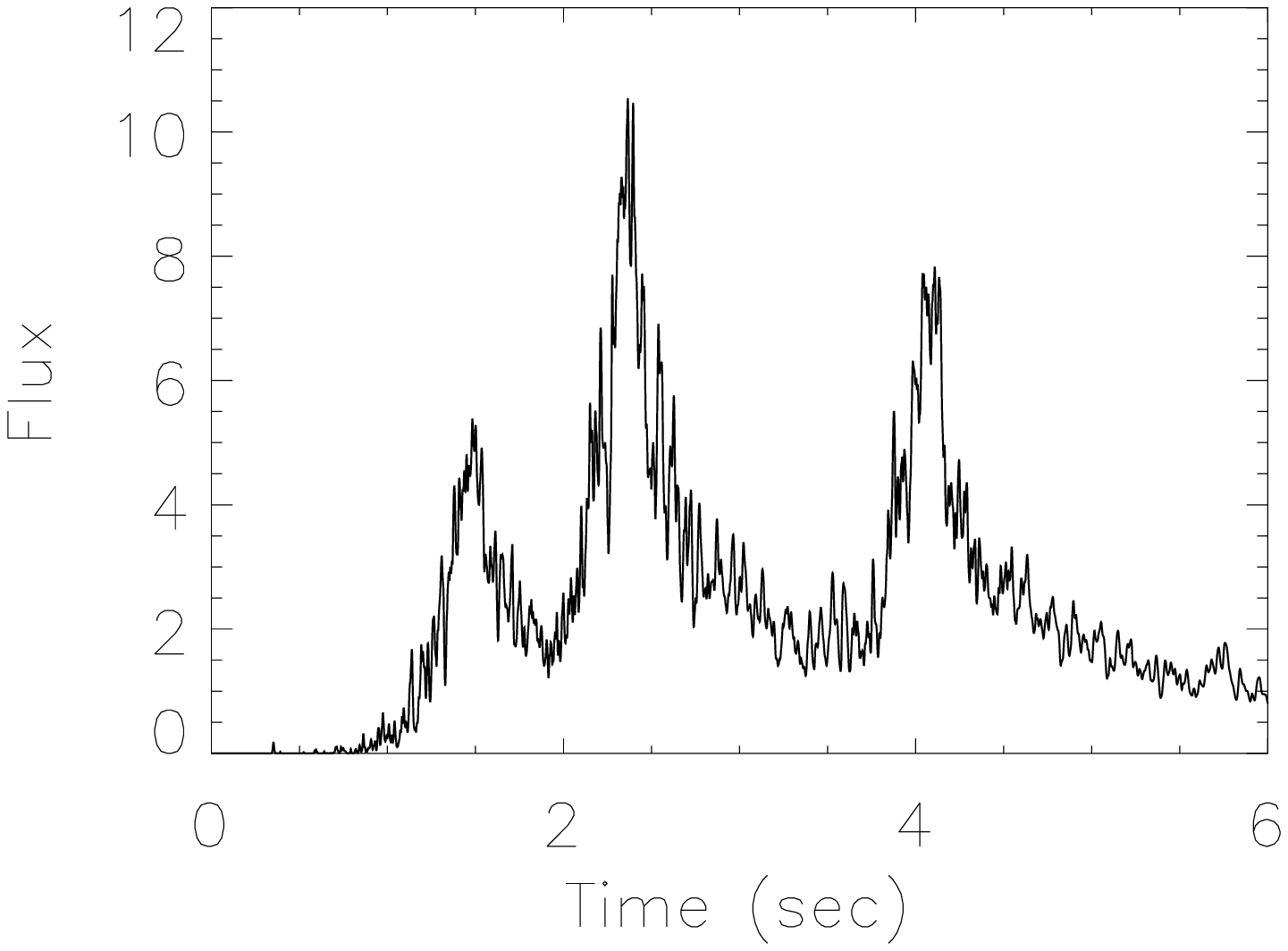}
\caption{\small{The Light curve of a GRB with three ICMART episodes. 
The parameters of each episode are close to those adopted in Fig.1b.}}\label{fig:GRB}
\end{center}
\end{figure}

As suggested by ZY11, a real GRB light curve may consist of
multiple ICMART events. In Figure \ref{fig:GRB}, we simulate three emission
episodes and superpose them together to make a mock GRB light curve.
We have varied $\Gamma_{\rm{ini}}$ and $\gamma_{\rm{ini}}$ around the values
$\sim 200$ and $\sim 3$, respectively, with small fluctuations in 
different episodes. Other 
parameters are the same as those adopted in Figure \ref{fig:direction} with a $45^\circ$
Gaussian $\phi$-distribution. The simulated light curve shows 
reasonable features as observed in some GRBs. 
We note that in reality the parameters
of different ICMART events could be more different, so that a
variety of light curves could be made, which may account for
the diverse prompt emission light curves as observed.

\newpage

\subsection{Power Density Spectrum (PDS) Analysis}

\begin{figure}
\begin{center}
\includegraphics*[width=4cm]{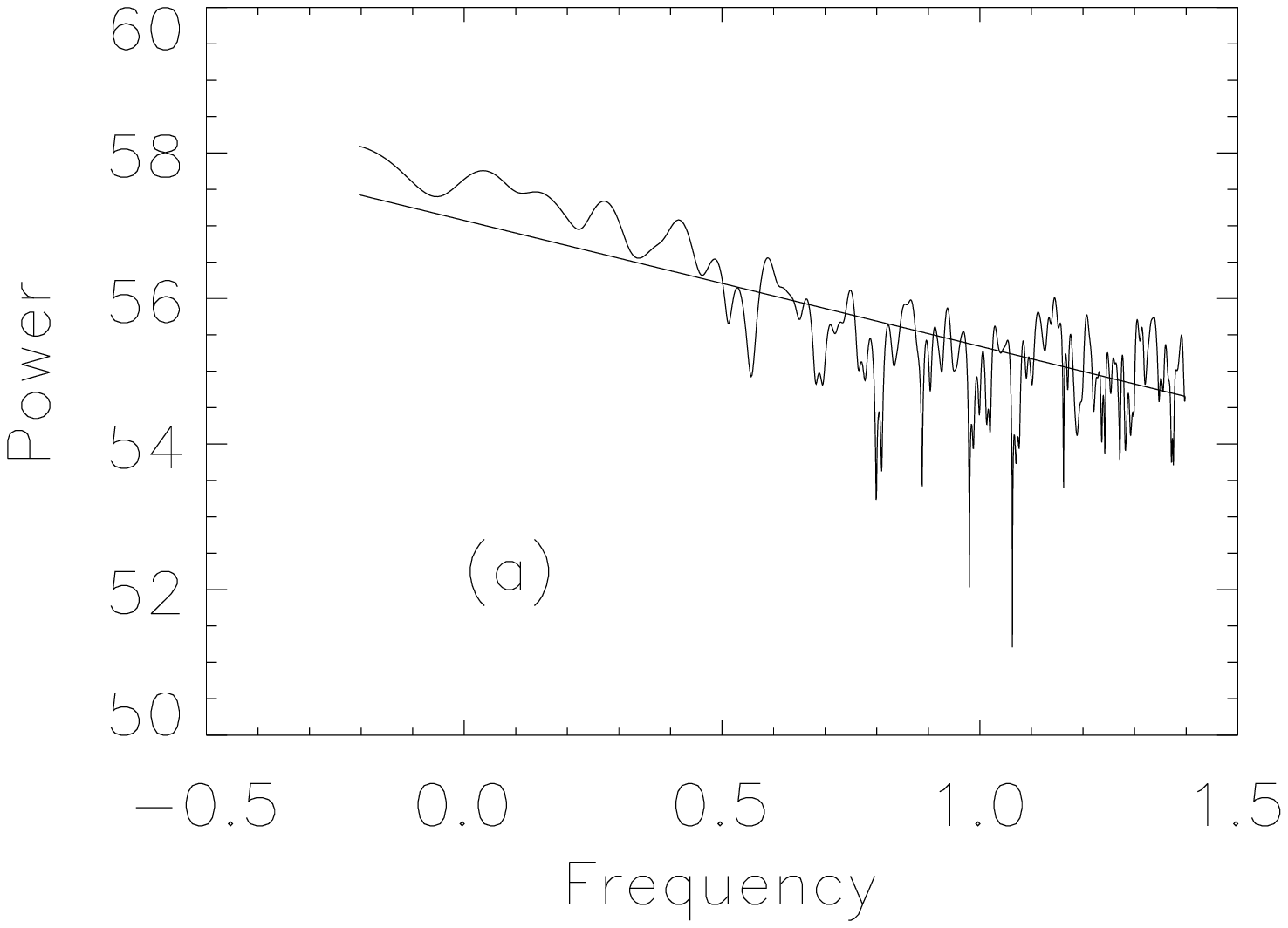}
\includegraphics*[width=4cm]{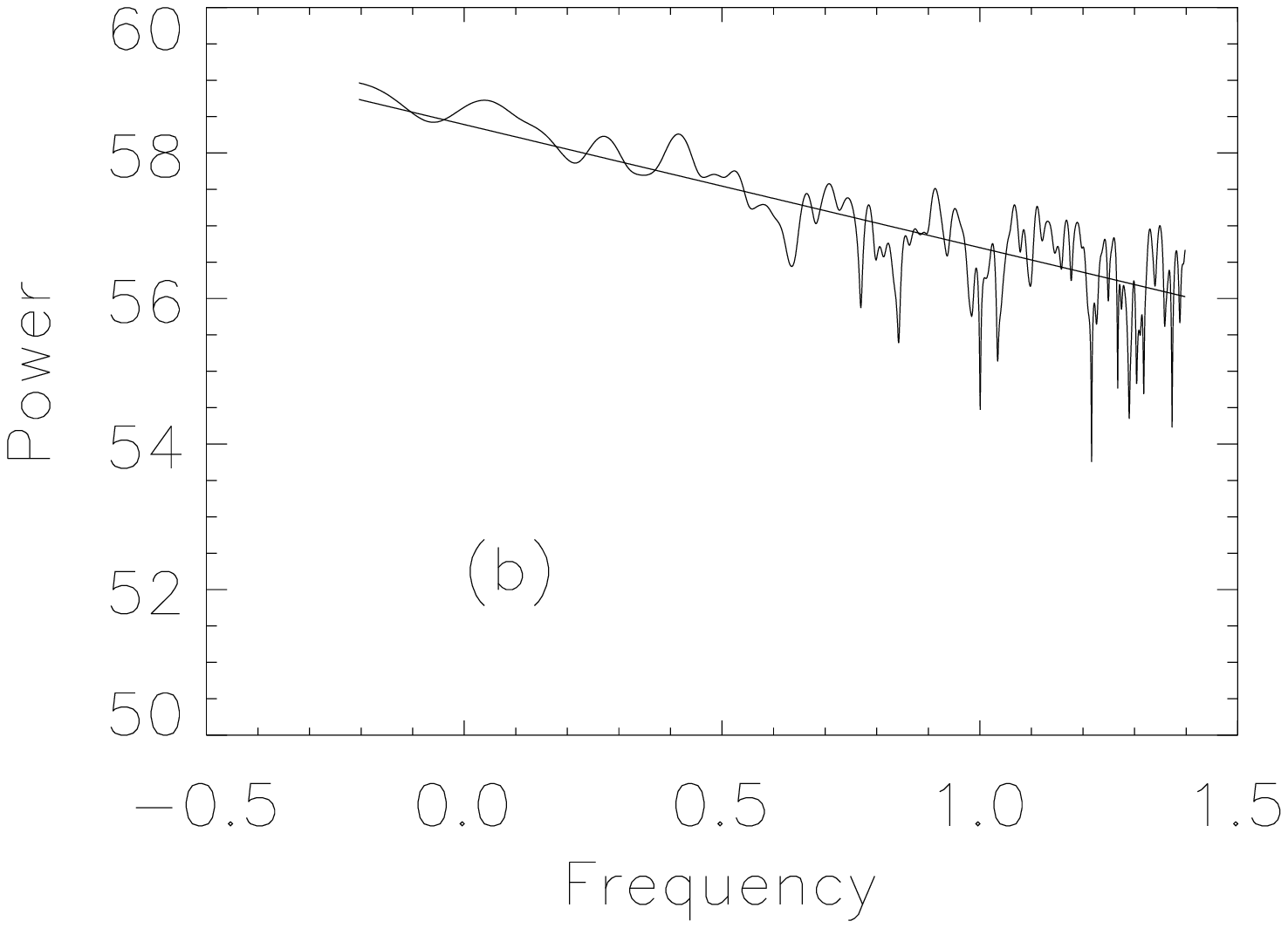}
\\
\includegraphics*[width=4cm]{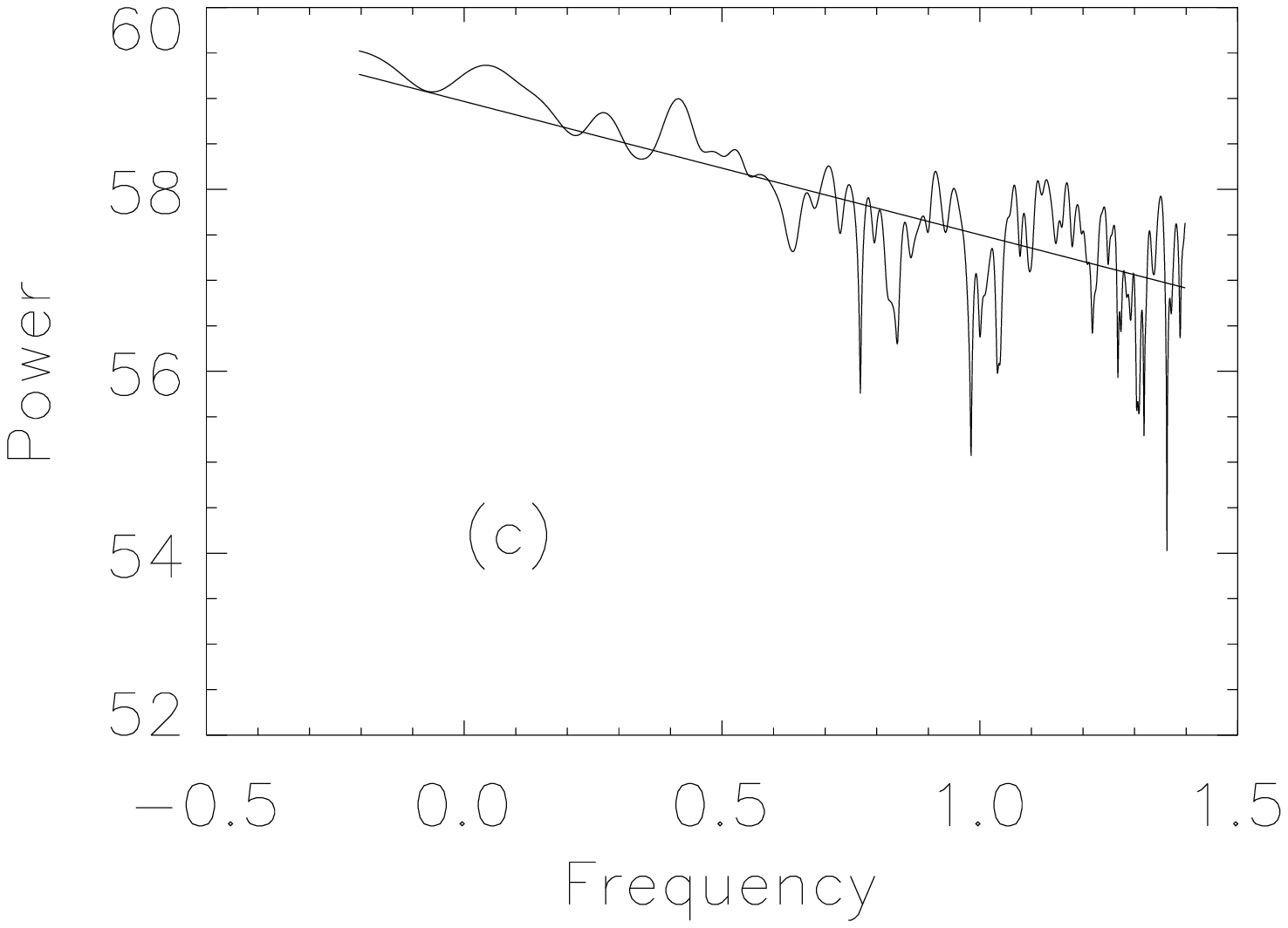}
\includegraphics*[width=4cm]{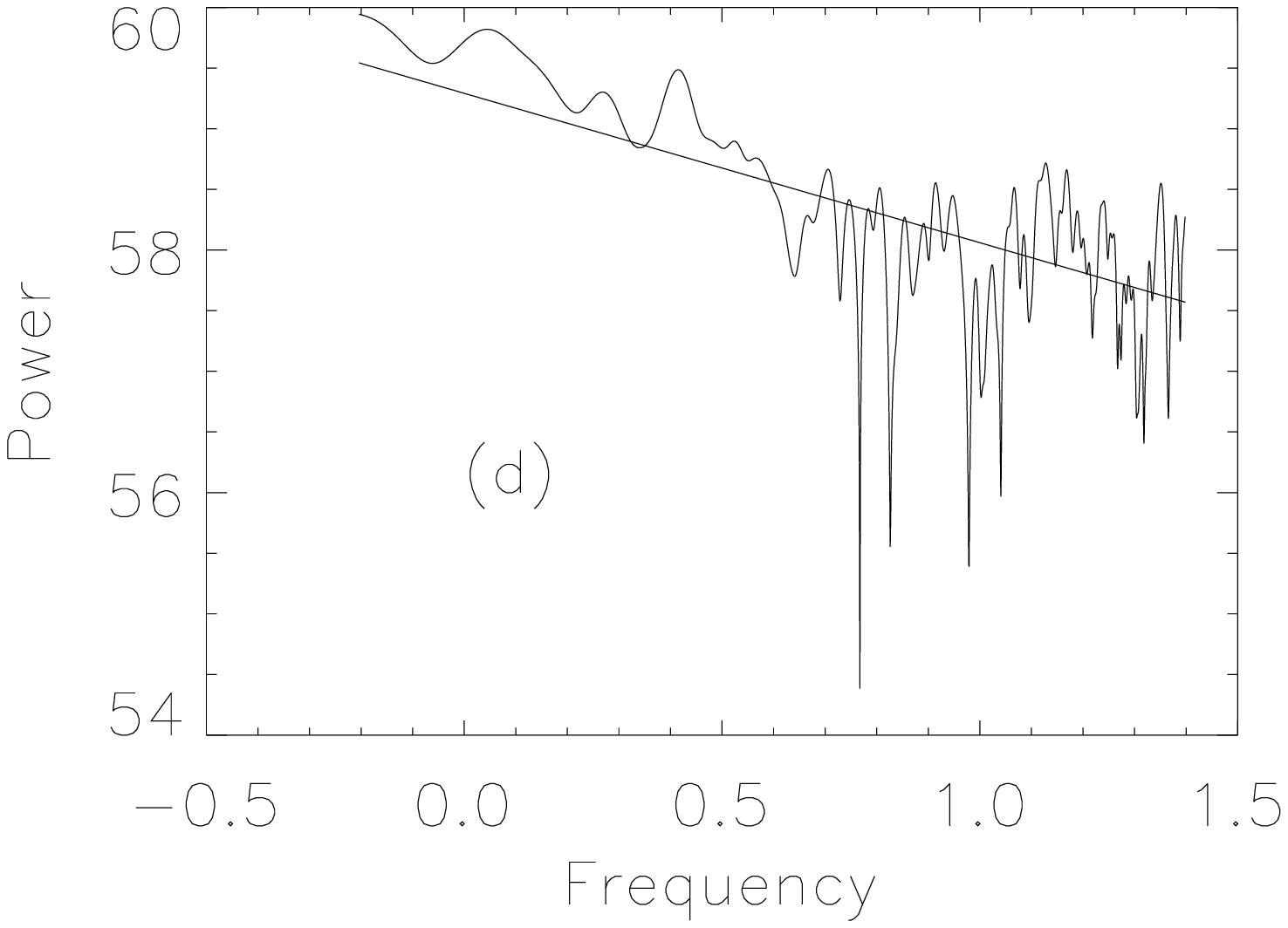}
\caption{\small{Power density spectrum (PDS) of sample light curves.
A $\phi$-Gaussian distribution with typical angle $45^\circ$ has been
adopted. The parameters are:  
top left: $\Gamma_{\rm{ini}} = 200$, $\gamma_{\rm{ini}}=3$, with power law 
index $p = -1.73$; top right: $\Gamma_{\rm{ini}} = 200$, $\gamma_{\rm{ini}}=8$, 
$p = -1.48$; bottom left: $\Gamma_{\rm{ini}} = 200$, $\gamma_{\rm{ini}}=14$, 
$p = -1.31$; bottom right: $\Gamma_{\rm{ini}} = 200$, $\gamma_{\rm{ini}}=20$, 
$p = -1.20$. Other parameters are the same as Figure 1(a). Note that these
values differ from those in Table 1, since the values in Table 1 are the averaged PDS slopes 
of 10 light curves.
}}\label{fig:PDS}
\end{center}
\end{figure}

In order to test whether our simulated light curves mimic the observed
ones, we also perform a PDS analysis of our results. 
In order to get robust PDS slopes, for each set of parameters, we perform 10
different Monte Carlo simulations to get 10 different light curves, derive
the PDS slope of each light curve, and calculate the average slope to stand
for this particular set of parameters. 
Some examples of PDSs are presented in Figure \ref{fig:PDS}.
Generally, the PDSs can be fit with a power law, with indices generally
steeper than $-1.8$. The averaged PDS indices for all the cases corresponding to
Figures 1, 2 , and 4-9 are collected in Table 1.
Observationally the PDS slopes are steeper in softer bands (e.g. \textit{Swift};
Guidorzi et al. 2012) than harder bands (e.g. BATSE; Beloborodov et al. 2000).
Our simulations recover this trend. The presented PDS values are taken
from the \textit{Swift} band. It is encouraging to see that the simulated values are
generally consistent with the \textit{Swift} data (Guidorzi et al. 2012). Our
simulations also show a turnover of PDSs in the high-frequency regime with a 
steeper index. Such a feature is seen in some GRBs.

\begin{table}
\caption{PDS Slopes of Simulated Light Curves}
\begin{center}
\begin{tabular}{cc}
\hline\hline
Figure & Slope \\
\hline
1(a) & -1.84\\
1(b) & -1.83\\
1(c) & -1.78\\
2(a) & -1.41\\
2(b) & -1.20\\
2(c) & -1.11\\
4(a) & -1.16\\
4(b) & -1.83\\
4(c) & -2.15\\
4(d) & -2.64\\
5(a) & -0.93\\
5(b) & -1.76\\
5(c) & -3.37\\
6(a) & -2.15\\
6(b) & -2.36\\
6(c) & -3.23\\
7(a) & -1.83\\
7(b) & -2.35\\
7(c) & -2.61\\
8(a) & -1.83\\
8(b) & -2.18\\ 
9(a) & -1.83\\
9(b) & -1.71\\
9(c) & -1.78\\
\hline
\end{tabular}
\end{center}
\end{table}

From Table 1, one can see that various parameters can affect the slope of a PDS.
Generally speaking, spikier light curves have more power in high frequencies
and therefore have a shallower PDS slope. Most PDS indices listed in Table 1
can be understood this way. For Figure \ref{fig:direction}, it is seen that more isotropic
distributions give steeper slopes. This is because the more isotropic cases
give more mini-jets contributing to the broad component, and thus enhance
the low-frequency power. Similarly, as shown in Figure \ref{fig:N}, 
a smaller number $N$ gives richer spiky features, and therefore gives a shallower PDS slope.
The $R$-dependence (Figure \ref{fig:R}) can be understood as the following: 
a larger $R$ corresponds to a longer curvature decay tail, on top of which 
rapid variability can be observed, so that the PDS slope is shallower. 
For the size effect (Figure \ref{fig:size1}), a smaller $L'$ can give rise to 
pulses with shorter duration and hence, a more dominant high-frequency power
and shallower PDS (Figure \ref{fig:size1}). When both $R$ and $L'$ co-vary, this
effect is still relevant, but somewhat compensated by the $R$ effect 
(Figure \ref{fig:size2}). Next, without a size distribution,
the PDS is steep (Figure \ref{fig:size-distribution}(b)). 
By introducing a size distribution, one has more contributions to short-time
variability from smaller sizes, so the PDS becomes shallower.
Finally, the light curves in a higher energy band are somewhat spikier
(Figure \ref{fig:bands}) and hence have a shallower PDS. This is consistent with
the finding of Guidorzi et al. (2012) and Beloborodov et al. (2000): 
using the \textit{Swift} BAT data, Guidorzi et al. (2012) obtained a steeper 
PDS slope than Beloborodov et al. (2000), who used the BATSE data 
(higher energy band) to perform the analysis.

\begin{figure}
\begin{center}

\includegraphics*[width=8cm]{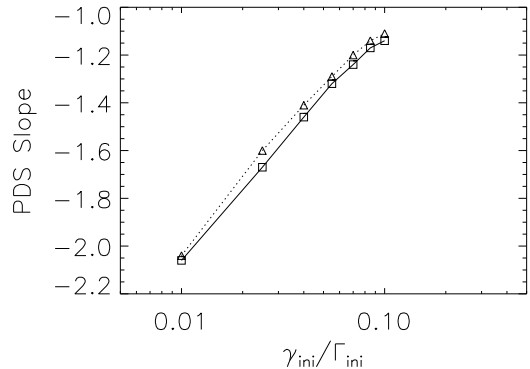}
\caption{\small{The PDS slope for different Lorentz factor contrast 
with a fixed $\Gamma=200$ . The triangles represent the PDS slopes 
of sample light curves without introducing parameter evolution. The 
dotted line connects the data points and gives the general trend of
dependence. The squares and the solid line represent the PDS slopes
as a function of Lorentz factor contrast when parameter evolution is
taken into account.}}\label{fig:PDS1}
\end{center}
\end{figure}

It is interesting to investigate the change of PDS slope due to the
change of the initial Lorentz factor contrast. As shown in Figure \ref{fig:evolution},
in principle one can have strong parameter evolution during one
ICMART event, which causes complicated evolution of the PDS behavior.
To avoid such strong evolution, we first fix $\Gamma_{\rm{ini}} = 200$,
and vary $\gamma_{\rm{ini}}$ so that the ratio $\gamma_{\rm{ini}}/\Gamma_{\rm{ini}}$
evolves in the range of $0.01 - 0.1$. In Figure \ref{fig:PDS1}, we present the PDS slope
as a function of $\gamma_{\rm{ini}}/\Gamma_{\rm{ini}}$. The triangles (and dotted
line) are calculated by turning off parameter evolution (i.e., keeping $\gamma$
and $\Gamma$ unchanged throughout), and the squares (and solid line)
are calculated by turning on the parameter evolution (Figure \ref{fig:evolution}). One can
see that the PDS slope becomes progressively shallower as $\gamma_{\rm{ini}}
/\Gamma_{\rm{ini}}$ increases. This is understandable, since a larger
$\gamma_{\rm{ini}}$ corresponds to a stronger fast emission component,
and therefore the light curves are spikier (see Figure \ref{fig:gamma-ratio}). One can 
tentatively draw the conclusion that a more magnetized outflow tends
to make spikier light curves. 

\begin{figure}
\begin{center}
\includegraphics*[width=8cm]{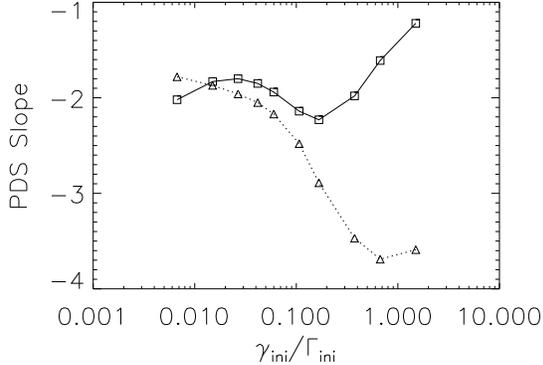}
\caption{\small{The PDS slope as a function of $\gamma_{\rm{ini}}/\Gamma_{\rm{ini}}$ 
for a fixed value of $\Gamma_{\rm{ini}} \gamma_{\rm{ini}} = 600$. 
The triangles represent the PDS slopes 
of sample light curves without introducing parameter evolution. The 
dotted line connects the data points and gives the general trend of
dependence. The squares and the solid line represent the PDS slopes
as a function of Lorentz factor contrast when parameter evolution is
taken into account.}}\label{fig:PDS2}
\end{center}
\end{figure}

Since the final Lorentz factor of the ejecta at the deceleration time
is proportional to $\Gamma_{\rm{ini}}\gamma_{\rm{ini}}$, and since observationally
the Lorentz factor at the onset of afterglow does not have a wide 
distribution (e.g., Liang et al. 2010), it is interesting to investigate
how the PDS slope depends on the Lorentz factor contrast when
$\Gamma_{\rm{ini}}\gamma_{\rm{ini}}$ is set to constant. In Figure \ref{fig:PDS2}, we present
the case of $\Gamma_{\rm{ini}} \gamma_{\rm{ini}} = 600$ for cases both without
and with parameter evolution. The range of the contrast is set to
$\gamma_{\rm{ini}}/\Gamma_{\rm{ini}} = 1/150$ (i.e. $\Gamma_{\rm{ini}} = 300$, 
$\gamma_{\rm{ini}} = 2$) to $\gamma_{\rm{ini}}/\Gamma_{\rm{ini}} = 1.5$
(i.e., $\Gamma = 20, \gamma = 30$). The convention is the same as
Figure 10. The dependence shows more complicated patterns. For the
case without evolution (triangles and dotted line), in general one
can see decrease of PDS slope when $\gamma_{\rm{ini}}/\Gamma_{\rm{ini}}$ 
increases (except the slight tilt at very large 
$\gamma_{\rm{ini}}/\Gamma_{\rm{ini}}$). This can be understood in the following
way. As $\gamma_{\rm{ini}}/\Gamma_{\rm{ini}}$ increases, one has two competing
effects. The increase of $\gamma_{\rm{ini}}$ tends to enhance the small
timescale variability. On the other hand, the decrease of $\Gamma$
tends to enlarge the $1/\Gamma$ cone, so that many more mini-jets
not beaming toward the observer could contribute to the slow component.
The net result after competition is that the latter effect wins, so
that the long-time variability is more enhanced, and hence, a steeper
PDS is obtained. This trend is overturned when $\gamma_{\rm{ini}}$ exceeds
$\Gamma_{\rm{ini}}$ near the end of the curve. 

When evolution is taken into account (squares and solid line), the 
situation is even more complicated. When $\gamma_{\rm{ini}}$ is small 
enough, the above-mentioned trend is retained. However, when $\gamma_{\rm{ini}}$
becomes large enough, evolution of $\gamma$ and $\Gamma$ becomes
significant (Figure \ref{fig:evolution}), so that quickly one can reach a 
regime with small $\gamma$ and large $\Gamma$. 
The average PDS would be dominated by this late phase, so that 
the general trend is reversed from the no-evolution case. 
In reality, since a real GRB light curve would usually be the 
superposition of multiple
ICMART events, the clean evolution expected in a
single ICMART event would be smeared out.


\section{Summary and Discussions}

In this paper we have simulated a sample of GRB prompt emission
light curves and PDSs
within the framework of the ICMART model (ZY11). This
model was developed to model GRBs whose jet composition is still
somewhat Poynting flux dominated in the emission region.
This was motivated by the non-detection of the photosphere
component in some GRBs (Zhang \& Pe'er 2009; Zhang et al. 2011).
Since the emission region has a moderately high $\sigma$, in
order to generate a reconnection/turbulence cascade envisaged by ZY11,
the energy dissipation region must have many locally Lorentz-boosted
emission regions, or mini-jets. The detected emission would be the
superposition of emissions from all these mini-jets, which beam to
random directions in the bulk comoving frame. 
Other global magnetic dissipation models for GRB prompt emission
have been proposed in the literature (e.g., Lyutikov \& Blandford 2003;
Giannios \& Spruit 2006). If these models invoke
runaway generation of mini-jets at a relatively large emission radius,
then the simulations in this paper also apply to those scenarios.

Lacking detailed numerical simulations for a reconnection/turbulence
cascade, we carried out a Monte Carlo simulation by
inputting many mini-jets with certain directional and temporal 
distributions within the ICMART scenario. We investigated the roles
of the directional distribution, Lorentz factor contrast, number of reconnection regions,
emission radius, size of the mini-jet, mini-jet size distribution, energy dependence, etc., 
in defining the light curves and their PDSs. 
We adopt our simulation parameters according to
observations (e.g., typical length of reconnection region $L^{'} = 5 \times
10^{11}$ cm corresponding to observed variability timescale $t_0 \sim 0.1$ s, 15 - 150 keV 
band for simulated light curves corresponding to \textit{Swift} BAT band, and so on), as well as
the requirements of the ICMART model itseft (e.g., emission region radius $R = 5 \times 10^{15}$ cm
in order to make sure that runaway reconnection can happen, and exponential growth of the number of
reconnection events with time). Within the ICMART framework, most of our parameters are 
physically related to each other self-consistently.

Even though some simplified assumptions are introduced so that the light 
curves may not fully represent the complex physics in an ICMART event,
our simulated light curves nonetheless show some encouraging
features that are consistent with the GRB prompt emission data.
The most noticeable feature is the superposition of an underlying
slow component and more erratic fast component, which seems to be
consistent with the data (Gao et al. 2012; Vetere et al. 2006).
The slow component is caused by the superposition of emission from
all the mini-jets in the emission region, while the fast component
is related to those mini-jets that happen to beam toward the
observer. We follow the physics of an ICMART event, including
the exponential growth of the reconnection region, dissipation of
the magnetic field energy (so that $\sigma$ drops with time), and
acceleration of the bulk ejecta during the energy dissipation
process and find that the erratic GRB light curves as observed
can be generally reproduced within the model. Among all the 
model parameters, the Lorentz factor contrast and the number of 
mini-jets play an important
role in defining the ``spikiness'' of the light curve. 
We also derived the PDS slopes of the simulated light curves, and
found that they are generally consistent with the data.
Generally speaking, the larger the contrast $\gamma_{\rm{ini}}/\Gamma_{\rm{ini}}$ 
(keeping $\Gamma_{\rm{ini}}$ constant), the shallower the PDS slope. 

Besides GRBs, the ``jet-in-the-jet'' scenario has been discussed 
in other astrophysical contexts. Giannios et al. (2010) interpreted
the fast TeV variability of active galactic nuleus jets using the mini-jet scenario.
Yuan et al. (2011) applied the scenario to account for the
gamma-ray flares of the Crab Nebula. Compared with earlier
work of Narayan \& Kumar (2009), Kumar \& Narayan (2009), and
Lazar et al. (2009), 
the new ingredient introduced in our paper is the exponential
growth of the number of mini-jets as a function of time,
as envisaged in the ICMART model (ZY11; see also Stern \& Svensson 1996). 
As a result, our model allows many mini-jets emitting simultaneously 
at any instant. This is the key ingredient to define the broad
component of each ICMART event.\footnote{In contrast, the previous
relativistic turbulence models (Narayan \& Kumar 2009; Lazar et al.
2009) only introduced a filling factor in time, so that they rarely
have multiple mini-jets emitting at any instant. So their simulated
light curves are too spiky, and do not have a slow component.}
A GRB light curve is composed of multiple ICMART events (Figure 8),
which are controlled by the erratic central engine activity.

In order to set up the Monte Carlo simulations, we had to
introduce a number of assumptions. These
include power-law distribution of the size of reconnection
regions, Gaussian shape of each pulse, same intrinsic radiation
spectrum for all emitters, exponential growth of
numbers of pulses with time, isotropic or Gaussian distribution
of the mini-jet directions, and so on. Some factors are 
still missing. For example, in the comoving frame of 
the jet bulk but outside the mini-jets, there would also be
particles that give rise to radiation. The effects of this
inter-mini-jet emission should be investigated  (e.g., Lin et al. 2013).

The physical conditions of real GRBs must be more complex than what
is modeled here, so that one may not
reproduce the full observational features of GRBs with the
simulations presented in this paper. 
Nonetheless, our simulations show the encouraging
results that the simulated light curves based on these simplified
assumptions can indeed reproduce some key features of the observations,
e.g. the slow and fast variability components and a variety of degree
of spikiness of the light curves. By changing parameters (e.g.,
$\phi$-distribution, Lorentz factor contrast, jet opening angle), 
diverse light curves can be generated, ranging from relatively 
smooth to relatively spiky ones. The PDSs of the simulated light 
curves are also generally consistent with the data.
All these suggest that the ICMART model may
be a good candidate to interpret GRB prompt emission. 

Within the ICMART theoretical framework, the following constraints can be
made to the model parameters. (1) To reproduce the general fast-rising
slower decay shape of broad pulses, the emission radius 
should be relatively large ($\sim 10^{15}$ cm and beyond).
(2) Since many GRBs show high-amplitude rapid variability, the GRB 
initial magnetization parameter $\sigma_{\rm{ini}}$ in the
emission region could be high (e.g., from several to hundreds). (3) The 
observed minimum variability timescale constrains that $L'$ cannot 
be too large and has to be $\leq 5\times 10^{11}$ cm. (4) In order
not to smear these peaks by overgenerating mini-jets, one also requires
a filling factor $f \ll 1$, suggesting that in these cases the global
$\sigma$ of the outflow after the ICMART event may not drop to unity.
(5) Erratic light curves with multiple episodes suggest that the GRB
central engine acts multiple times to eject highly magnetized shells
so that multiple ICMART events can be generated within one burst.
(6) The existence of smooth-pulse GRBs suggests that in some cases the
$\sigma_{\rm{ini}}$ is not much larger than unity (so that $\gamma_{\rm{ini}}$
is not much larger than unity), or there are so many mini-jets operating
simultaneously. Other information (e.g. polarization properties and
prompt emission efficiency) is needed to break the degeneracy.

\acknowledgments We thank Kohta Murase, Pawan Kumar, Zi-Gao Dai, He Gao, 
Da-Bin Lin, and Chun Li for useful discussion, and an anonymous referee for very helpful
suggestions. This work is partially
supported by NSF through grant AST-0908362. Bo Zhang 
acknowledges a scholarship from China Scholarship Council
for support.

\end{CJK*}


\end{document}